 \documentclass[aps,pra,letterpaper,superscriptaddress,twocolumn,showpacs,floatfix,10pt]{revtex4-1}

\usepackage{amsfonts}
\usepackage{amsmath}
\usepackage{amssymb}
\usepackage{graphicx}
\usepackage{epstopdf}
\usepackage{epsfig}
\usepackage{color}
\usepackage{mathtools}

\usepackage{hyperref}

\begin{document}

\title{Transition of a $\mathbb{Z}_3$ topologically ordered phase to  trivial and critical phases  }

\author{Ching-Yu Huang}
 \affiliation{C. N. Yang Institute for Theoretical Physics and Department of Physics and Astronomy, State University of New York at Stony Brook, NY 11794-3840, United States}  
 
\author{Tzu-Chieh Wei}
 \affiliation{C. N. Yang Institute for Theoretical Physics and Department of Physics and Astronomy, State University of New York at Stony Brook, NY 11794-3840, United States}  

\vfill
\begin{abstract}
Topologically ordered quantum systems have robust physical properties, such as quasiparticle statistics and ground-state degeneracy, which do not depend on the microscopic details of the Hamiltonian. 
We consider topological phase transitions under a deformation such as an effective string tension  on a $\mathbb{Z}_3$ topological state.  
This is studied  in terms of the gauge-symmetry preserved quantum state renormalization group, first  proposed by He, Moradi and Wen [Phys. Rev. B {\bf 90}, 205114 (2014)].  
In this approach  modular matrices $S$ and $T$ can be obtained and used as order parameters to characterize the topological properties of the phase and determine phase transitions. 
From a mapping to a classical 2D Potts model on the  square lattice, the critical string tension, at which the transition to a topologically trivial phase takes place, can be obtained analytically and agrees with the numerically determined value. 
Such a transition can be generalized to a $\mathbb{Z}_N$ topological model under a string tension and determined in the same way.
With different deformations,  the $\mathbb{Z}_3$ topological phase can also be driven to a critical phase which contains, in the large deformation limit, the wavefunction analogous to the Rokhsar-Kivelson point in the quantum dimer model in one case and the fully packed loop model in another case. 
\end{abstract}

\maketitle

\section{ introduction}

Conventional phases and their transitions  in condensed-matter systems can be  understood by Landau symmetry breaking~\cite{arodz2003patterns}.
However, new phases have emerged over the past few decades such as  integer quantum hall~\cite{QHE80, Laughlin81} and fractional quantum hall~\cite{FQHE1982} effects  have evaded the usual Landau-Ginzburg-Wilson paradigm.
More recently, microscopic spin models exhibiting topological orders were also constructed, such as the toric code, quantum double models and string-nets~\cite{Dennis2002, Kitaev2003, string_wen03, string_wen05}. 
These new quantum phases are called topological phases and they cannot be characterized by a local order parameter. 
Instead, they are found to possess the so-called topological order~\cite{Wen1989, Wen1990} characterized by properties such as the ground-state degeneracy, nontrivial quasiparticle statistics~\cite{Wen_TO1990, Wen_TO1993,Kitaev20062, Bais_2012, Zhang2012}, and more recently nonzero topological entanglement entropy~\cite{Kitaev_2006,  Levin_2006}.

Recent progress  on the general scheme of classifying these topological orders is also vitalized by using the notation of entanglement~\cite{Pollmann_2010,Pollmann_2011,Xie_TPS_2010, Xie_LU_2010, Haldane_2008}. 
Specifically, intrinsic topological orders have patterns of long-range entanglement that cannot be achieved via local unitary transformation from a trivial product state.  
The local unitary transformations can only remove  short-range entanglement, i.e., quantum correlations between neighboring sites. 
Due to its robustness against the local operations, the long-range entanglement gives rise to topological order of the nontrivial ground state.
Long-range entanglement is manifested in the so-called topological entanglement entropy~\cite{Kitaev_2006,  Levin_2006,TEE_0701,TEE_0702, TEE_08, TEE_11,TEE_12,Vidal2013, TEE_PEPS_cirac11}  and other related entanglement quantities~ \cite{negativity201301, negativity201302,orus14, Multipartite_2014}.

However,  in general, it is still numerically challenging to access large system sizes to extrapolate accurately the topological entanglement  to determine whether a topological phase exists. 
Even if one manages to obtain nonzero topological entanglement entropy, there are possibly different topological phases that possess the same value of the entanglement.

Quasiparticle statistics, related to the generally non-abelian geometric phases, represented by the modular $S$ and $T$~\cite{Wen_STmatrix} matrices, provide a more refined characterization of a topological order.  $S$ and $T$ are representations of $90^\circ$-rotation and the Dehn twist, respectively, on a two-dimensional torus. 
These topological quantities  do not depend on the microscopic details of the Hamiltonian, and can be used as order parameters, albeit non-local.
It turns out that one can, given a complete basis states in the degenerate ground space, exploit entanglement, with respect to a bipartite cut, to deduce the set of  minimally entangled states (MESs). These states form a special complete set of bases~\cite{Grover_11,Zhang2012, Zhu13,Vidal2013,Tu_2013,He14_dmrg} in the ground space, and from them the modular matrices $S$ and $T$ can actually be deduced.  
Methods such as Quantum Monte Carlo and the exact diagonalization have been used in such computations~\cite{Zhu13, Siddhardh14, MES_GM2014}.

The recent development  of matrix product states (MPS)~\cite{ mps_1995,mps_2003,mps_2004} and the 2D tensor product states (TPS)~\cite{ peps_2004, Murg2007,Vidal2007, Vidal2008,Levin_TRG2007,Xiang_08}, which are the generalization of the density-matrix renormalization group (DMRG) method~\cite{dmrg1992, dmrg1993} has given rise to alternative approaches to obtain ground states and thus MESs.
These tensor-network or tensor-product states also were constructed as variational wavefunctions for optimizing Hamiltonians~\cite{trg08-Wen,Xiang_08, Vidal2007,  Vidal2008, Murg2007, Vidal2009}, but they have also been shown to exactly represent a large class of topological states,
 including both non-chiral~\cite{Verstraete_04, Verstraete_06,Oliver_09, Gu_09}  and chiral~\cite{Read_2013,Wahl_2013}  topological order.
The DMRG itself has also been used to obtain the MES systematically~\cite{Vidal2013,Zaletal14,He14_dmrg}.
In particular, via techniques introduced in Refs.~\cite{Vidal2013, Zhu13, Zaletal14, Siddhardh14, He_Wen_2014}, the MESs corresponding to different quasiparticle excitations can be obtained and  the modular matrices can be determined from MESs.
We shall follow the approach proposed by He, Moradi and Wen~\cite{He_Wen_2014},  who showed that the gauge-symmetry structure of TPS can give us information about  topological order. 
In particular, applying the gauge-symmetry preserved tensor renormalization group (GSPTRG) to $\mathbb{Z}_2$ topological orders under deformation of the wavefunction via a string tension, He et al. obtained the modular matrices as a topological order parameter to characterize the topological phases and their phase transition from the $\mathbb{Z}_2$ topologically ordered phase to  a trivial phase (that is adiabatically connected to a product state).  
Here, we employ the same approach and apply it to $\mathbb{Z}_N$ topological models under deformations and characterize the topological phase with the modular matrices and locate the phase transitions to topologically trivial phases. 
For one particular type of deformation, we can map the wavefunction norm square to the partition function of the classical $N$-state Potts model and obtain the analytic critical string tensions. 
The numerical results via GSPTRG agree very well with those via the mapping. 
The unexpected result we obtain is that under different deformations, the $\mathbb{Z}_3$ topologically ordered phase can also be driven to a critical phase, in addition to the trivial product-state phase. 
The phase diagrams of the $Z_3$ topological phase under different deformations considered in this paper are summarized in Fig.~\ref{phase_diagram}.

The paper is organized as follows. 
In Sec.~\ref{review} we review the notion of topological order and gauge symmetry preserved tensor renormalization group  which can be used to  identify intrinsic topological orders; 
In Sec.~\ref {results}  we discuss  how the deformation can be applied to the $\mathbb{Z}_N$ topologically ordered model via the tensor product states and how to use  GSPTRG to compute the modular matrices for characterizing the topological order. 
There, we also show a useful mapping form the  $\mathbb{Z}_N$ model under a string tension to  the two-dimensional classical $N$-state Potts model, from which the critical point between the topologically ordered phase and a trivial phase can be obtained analytically and compared with numerics. 
In Sec.~\ref {critical}, we evaluate the modular matrices and correlation function for the $\mathbb{Z}_3$ model and discuss the critical phases. 
We conclude in Sec.~\ref{conclusion}.

\section{Modular matrices and gauge symmetry preserved tensor renormalization group} \label{review}

\subsection {Modular matrices}

One of the exotic features of the topological order is the nontrivial quasiparticle statistics, which can be obtained by modular transformations on degenerate ground states on the torus, giving rise to the modular $S$ and $T$ matrices. 
This approach of characterizing topological orders has become quite fruitful recently~\cite{Zhang2012, Vidal2013, Michael_2013,He_Wen_2014}.
The modular matrices, or $S$ and $T$ matrices, are generated respectively by  the $90^{\circ}$ rotation and the Dehn twist on a set of degenerate ground states  on the torus. 
The elements of $S$ matrix express the mutual statistics of the quasiparticles, whereas 
the $T$ matrix express the twisting a quasiparticle wavefunction along an axes by $360^{\circ}$.

Specifically, to obtain the modular matrices, we need to first determine all the degenerate ground states  $\{  | \psi _a \rangle \}_{a=1}^{N}$ of the system.
The $S$ and $T$ matrices are defined as follows~\cite{ Wen1989,Wen1990,Levin_2006}:
\begin{align}
& \langle \psi_a | \hat{S} | \psi_b \rangle = e^{-\alpha_S V+\mathfrak{o} (1/V)} S_{ab}  \notag  \\
& \langle \psi_a | \hat{T} | \psi_b  \rangle= e^{-\alpha_T V+\mathfrak{o} (1/V)} T_{ab},
\end{align} 
where $\hat{S} $ and $\hat{T} $ are the transformations of the $90^{\circ}$ rotation and the Dehn twist respectively on a torus with lattice size $V$, $\alpha_S$ and  $\alpha_T$  are non-universal constants, and $S_{ab} $ and $T_{ab} $ are elements of the modular matrices. 
The information of quasiparticles statistics and their fusion rule are encoded in the $S$ and $T$ matrices~\cite{Verlinde_1987, Liu_2013,Wen_STmatrix,Moradi2014}.

In particular,  from the gauge structure of tensor product states at the fixed point (TPS)~\cite{Xie_TPS_2010,Swingle2010,Schuch2010,Oliver2014}, the degenerate ground states can be obtained by inserting the gauge transformation to TPS, i.e., operating on the bond or virtual degrees of freedom by the appropriate gauge transformation. 
The degenerate ground states can be labeled as $|\psi(g,h)\rangle$ with gauge transformations $( g,h) $ applied on the internal indices along two directions. 

Next, we describe the real-space renormalization group approach and how the modular matrices can be obtained.

\subsection { Gauge symmetry preserved tensor renormalization group (GSPTRG) }

An symmetry preserved tensor renormalization group procedure exist for quantum states based on the tensor product representation
\begin{align}
|\psi\rangle=\sum_{s_1,s_2,... s_m...}\text{tTr}(A^{s_1}A^{s_2}...A^{s_m}...)|s_1 s_2... s_m...\rangle, 
\label{TPS}
\end{align} 
where $A^s_{\alpha,\beta,\gamma,...}$ is a local tensor with physical index $s$ and internal indices $\alpha\beta\gamma$ etc. $\text{tTr}$ denotes tensor contraction of all the connected inner indices according to the underlying lattice structure.   
The norm of TPS is given by 
\begin{align}
\langle \psi |  \psi \rangle =\text{tTr} (\mathbb{T}^1  \mathbb{T}^2  \mathbb{T}^3... \mathbb{T}^m...   ), 
\label {ST_TPS}
\end{align} 
where the local {\it double tensor\/} $\mathbb{T}^i$ can be formed by merging two layers tensors $A$ and $A^*$ with the physical indices contracted,
\begin{equation}
\mathbb{T}\equiv \sum _s  (A^s_{\alpha,\beta, \gamma,\delta... })  \times  (A^s_{\alpha',\beta', \gamma',\delta'... }) ^*
\end{equation}
However, it is generally computionally hard to calculate exactly the tensor trace (tTr) or the contraction of the whole tensor network in two and higher dimensions 
This imposes the hurdles of an exponentially hard calculation. 
Several approximation schemes have been proposed as solutions in this context, such as iPEPS~\cite{Vidal2009} algorithm, the corner transfer matrix method (CTMRG)~\cite{CTMRG1997}, and tensor renormalization approach~\cite{Levin_TRG2007}, all of which tackle this problem essentially by scaling the computational complexity down to the polynomial level for calculating the tensor trace.

We express  tensor renormalization group (TRG)  approach which is akin to the real space renormalization in the way that at each step, the RG is structured by merging sites (by contracting respective tensors) and truncating the bond dimension according to the relevance of the eigenvalues in the Schmidt decomposition of the old tensors.
Actually, by doing several steps of TRG, the double tensor will flow to a fixed point. The information of the ground states can be extracted from the fixed point tensor.

A many body wave function  and Hamiltonian may be invariant under certain transformations that correspond to symmetries. 
The symmetry group divides the Hilbert space of system into symmetry sectors that can be labeled by the quantum number or conserved particle number. 
Symmetries can be used to improve the numerical methods, such as used in the exact diagonalization method and density matrix renormalization group~\cite{ed_symmetry, dmrg_symmetry}, as well as matrix product states and tensor product states~\cite{U1TPS_singh10,SPQSRG_huang13,He_Wen_2014}.
The  TRG method has itself emerged as an alternative approach for dealing with quantum spin systems. 
However,   the local decomposition usually breaks the symmetry sectors if the symmetry is not strictly enforced. Then, it would flow to the wrong fixed point. 
In this section, we will express the gauge symmetry preserved tensor renormalization group (GSPTRG) procedure proposed by He, Moradi and Wen~\cite{He_Wen_2014}  for TPS and demonstrate its effectiveness in identifying the topological order from the fixed point tensors. 
GSPTRG differs from TRG only when we decompose the tensor. 
The matrix would has blocks corresponding to the symmetry sectors. 
Once we keep the block structure, the symmetry will be preserved.

\begin{figure}[ht]
\center{\epsfig{figure=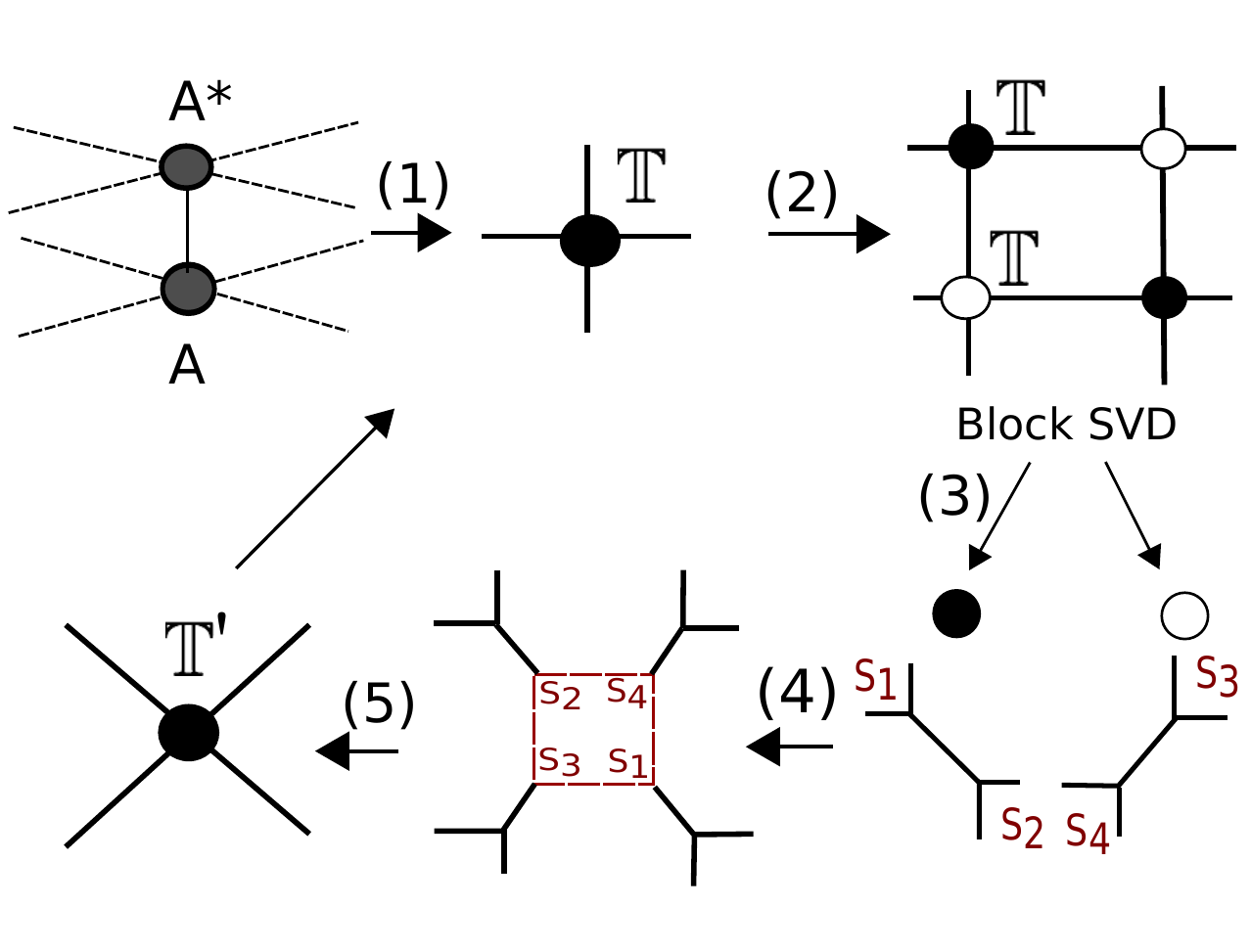,angle=0,width=8.5cm}}
\caption[] {Schematic procedure of the GSPTRG}  \label{SPTRG}
\end{figure}

In Ref.~\cite{He_Wen_2014}, the authors showed how to preserve the  $\mathbb{Z}_2$ gauge symmetry to determine the phase  diagram of  $\mathbb{Z}_2$ and double-semion model with string tension. 
The $\mathbb{Z}_2$ topological model can be realized in toric code model. 
The ground state of toric code model is an equal weight superposition of all closed string loops and it can be represented by tensor product state with virtual dimension $\chi=2$.  
When putting the Hamiltonian on the torus, it has four-fold degeneracy ground states which can be obtained by inserting the string operators,  such as  Pauli matrix $Z$, to the virtual bond degrees of freedom of one particular ground-state wavefunction. 

Below we shall consider the case of  $\mathbb{Z}_N$  topologically ordered phases for which the Hamiltonian is the generalized toric code model~\cite{Zn_2012,He_Wen_2014}, so the $\mathbb{Z}_N$ topological order has $\mathbb{Z}_N$ gauge symmetry. 
To implement GSPTRG, first we form a double tensor  $\mathbb{T}$ on each site $ \mathbb{T} = \sum _s  (A^s_{\alpha,\beta,\gamma,\delta })  \times  (A^s_{\alpha', \beta',\gamma',\delta' }) ^*   $ as shown in Fig.~\ref{SPTRG} (1). 
The double tensor  $\mathbb{T}$  will has $\mathbb{Z}_N  \times \mathbb{Z}_N$  gauge symmetry. 
Due to   such a gauge symmetry,  the elements of double tensor are non-zero only when ${\alpha +\beta +\gamma +\delta =0}$ (mod $N$) and  ${\alpha' +\beta' +\gamma' +\delta' =0}$ (mod $N$).

\begin{figure}[ht]
\center{\epsfig{figure=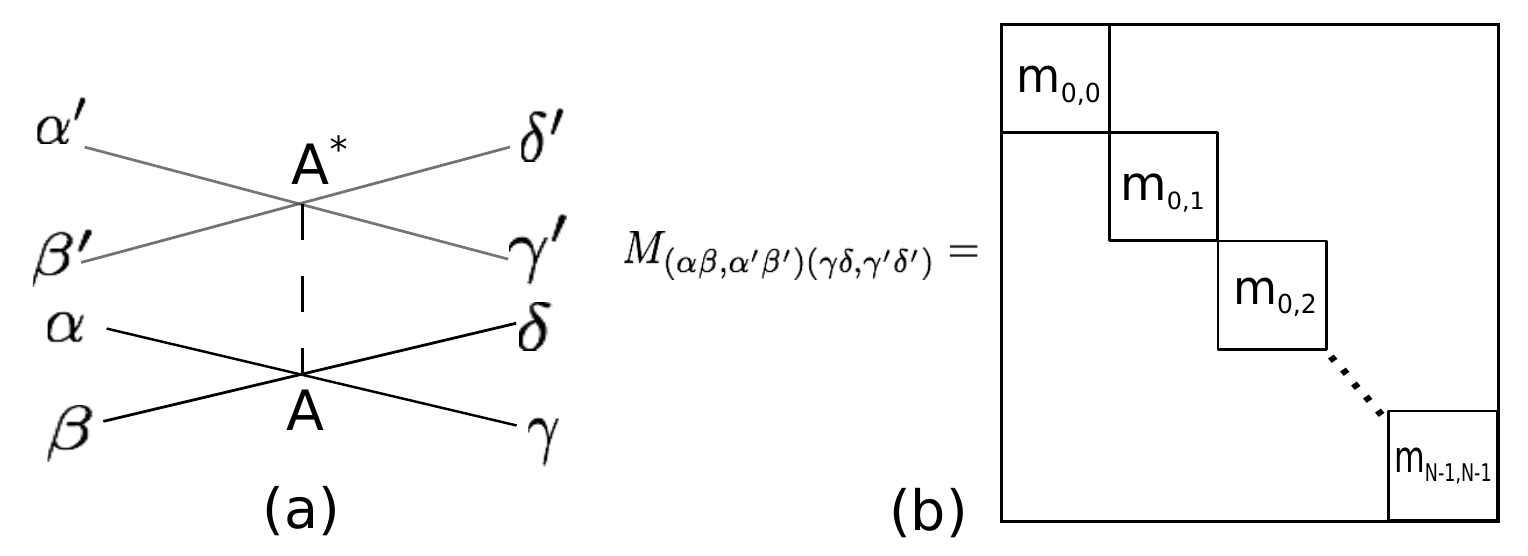,angle=0,width=8.5cm}}
\caption[] {(a) The double tensor structure. (b) The $M$ is  block diagonalized by the quantum number. } \label{Block_form}
\end{figure}

We then view the tensor $\mathbb{T}$ as a matrix $M_{\alpha \beta  \alpha' \beta',   \gamma \delta\gamma'  \delta' }  = \mathbb{T}_{\alpha\alpha',  \beta \beta',  \gamma\gamma',  \delta \delta'}  $.
The first step of the coarse graining is to decompose a rank-four tensor $M$ (e.g., the double tensor $\mathbb{T}$) into  two rank-three tensors. 
We do it in two different ways on black and white tensors (see Fig.~\ref{SPTRG} (3)). 
Due to  a such gauge symmetry, the tensor
\begin{align}
M_{\alpha \beta \alpha' \beta', \gamma \delta\gamma'  \delta' } = \bigoplus _{p,q=1} ^{N} m_{p,q}, 
\end{align} 
would be block diagonalized by the quantum number. For example, the each block $m_{p,q}$ obey the rule 
\begin{align}
&{\alpha+\beta =p}  \quad        \text{mod}  \quad      N \notag \\ 
 &{\gamma+\delta  =N-p } \quad     \text{ mod} \quad      N  \notag  \\
&{ \alpha'+\beta' =q } \quad     \text{ mod} \quad      N   \notag  \\
&{\gamma'+\delta'  =N-q } \quad   \text{mod}\quad      N, 
\end{align} 
where  $p,q=0,1,2,3,...,N-1$ as shown in Fig.~\ref{Block_form} (b). Then, singular-value decomposition (SVD) is performed in each block. 
 As mentioned above, the tensor contraction is an exponentially hard calculation. 
 Here,  a cutoff $\chi_c$ might be necessary on the dimension of double tensor to keep the computation efficient.
When making the truncation, we need to preserve the symmetry structure of the tensor by keeping the blocks together.
Such symmetry considerations apply similarly to all symmetry groups.

 After the decomposition, the lattice structure is changed. 
 The second step is to form a new rank-four tensor denoted by $\mathbb{T}'$  as shown in Fig.~\ref{SPTRG} (5).  
To do this, we combine the resultant four tensors that meet at s square to form a new tensor as shown in Fig.~\ref{SPTRG} (4). 
After doing several steps GSPTRG, the double tensor will gradually flow to a fixed point tensor that preserves the gauge symmetry.

The modular matrices can be evaluated and monitored during the process of the RG steps by performing three steps.
(i)  Inserting the gauge transformations $g,h,g',h'$  into  the internal indices $\alpha, \beta, \alpha', \beta',$ respectively of the fixed point double tensor as shown in Fig.~\ref {Block_form} (a) to determine   $ \big \langle  \psi(g',h') | \psi(g,h)  \big \rangle$.

(ii) Performing the rotation and the Dehn twist operators on the ground state wave function.
The operators performing on the physical indices can be achieved or replaced by appropriate gauge operations to internal indices,
\begin{align}
&   \big\langle  \psi(g',h') | \hat{T} | \psi(g,h)    \big\rangle =    \big\langle  \psi(g',h') | \psi(g,gh)       \big\rangle \notag \\
&   \big\langle  \psi(g',h') | \hat{S} | \psi(g,h)   \big\rangle =    \big\langle  \psi(g',h') | \psi(h,g^{-1})  \big\rangle \notag.
\end{align}

(iii) Finally, tracing all internal indices of the fixed point tensor.

For the  $\mathbb{Z}_2$ topological phase in Ref.  \cite{He_Wen_2014}, the  $\mathbb{Z}_2$ gauge symmetry is generated by  $\sigma ^z$ acting on each internal indices.  (We note that for different forms of ground-state wavefunctions, the gauge operator might be $\sigma^x$.)
The same rule holds in the  $\mathbb{Z}_N$ topological phase that generalizes from this case. 
The gauge symmetry can be generated by the $N \times N $ operator $Z$  at which all elements are zero  except diagonal term $Z_{k,k} = \exp { \frac{2\pi i (k-1)}{N}}; k =1,2,3,...N$. 

Instead of viewing the gauge operators $g,h$ applied to internal indices, we  can also understand  the degenerate ground state $ | \psi(g,h)  \big \rangle$ in terms of closed string operators $\hat{g},\hat{h}$ defined in the physical dimension. 
The $ \hat{S} $ and  $\hat{T}$ transformation can be written in the basis of the degenerate ground states $ | \psi(\hat{g},\hat{h})  \big \rangle$ generated by appropriate closed string operators  $\hat{g},\hat{h}$ on the torus along the vertical and horizontal directions, respectively. 
The string operators for the particular $\mathbb{Z}_N$ phase (i.e. $\mathbb{Z}_N$ toric code)  can be chosen as  $\mathcal{Z}^q \equiv  (Z )^q \otimes (Z^{\dagger})^ q \otimes     (Z )^q \otimes (Z^{\dagger})^ q   \dots $, $q=0,1,2,...,N-1$ and  $ \mathcal{Z}^0 = \mathcal{I}$.
The $ | \psi( \mathcal{I}, \mathcal{I})  \big \rangle  \equiv    | \psi \rangle $ is our reference ground state represented by the tensor product state in  Eq. \ref{ZN_TPS}.

Our goal is to obtain the modular $S$ and $T$ matrices.
We  can obtain the $N^2  \times N^2$ $S$-matrix generated by  $90^\circ$-rotation and it is  given as
$S_{a,b} =   \big\langle   \psi( \mathcal{Z}^{a_1},\mathcal{Z}^{a_2})  \big\rangle | \hat{S}   | \psi(\mathcal{Z}^{b_1},\mathcal{Z}^{b_2})  \big\rangle
                  = \big\langle   \psi(\mathcal{Z}^{a_1},\mathcal{Z}^{a_2}  \big\rangle   | \psi(\mathcal{Z}^{b_2},\mathcal{Z}^{-b_1})  \big\rangle  $;   where
                  $a_1,a_2,b_1,b_2 = {0,1,2,...N-1} $.  Both matrix indices $a$ and $b$ range from $1$ to $N^2$  where $a = a_1\, N+a_2+1$ and $b = b_1\, N+b_2+1$. 
Similarly, to apply  the Dehn twist to all degeneracy ground states we also can get the  $N^2  \times N^2$ $T$-matrix:  
$T_{a,b} =   \big\langle   \psi( \mathcal{Z}^{a_1},\mathcal{Z}^{a_2}) | \hat{T}   | \psi(\mathcal{Z}^{b_1},\mathcal{Z}^{b_2})  \big\rangle
                  =  \big\langle   \psi(\mathcal{Z}^{a_1},\mathcal{Z}^{a_2}   | \psi(\mathcal{Z}^{b_1},\mathcal{Z}^{b_1+b_2})  \big\rangle  $.
For example,   the  4-by-4   $T$-matrix for the  $\mathbb{Z}_2$ toric code is given  as follows:
\begin{widetext}
\begin{align} 
\label{Tmatrix}
T =  \left(
\begin{array}{ccccc}
   \big\langle \psi(\mathcal{I},\mathcal{I})   | \psi(\mathcal{I},\mathcal{I})  \big\rangle &\big \langle \psi(\mathcal{I},\mathcal{I})   | \psi(\mathcal{I},\mathcal{Z})  \big\rangle 
&\big\langle  \psi(\mathcal{I},\mathcal{I})   | \psi(\mathcal{Z},\mathcal{Z})  \big\rangle &\big\langle  \psi(\mathcal{I},\mathcal{I})   | \psi(\mathcal{Z},\mathcal{I}) \big\rangle \\
  \big\langle \psi(\mathcal{I},\mathcal{Z})   | \psi(\mathcal{I},\mathcal{I})  \big\rangle & \big\langle \psi(\mathcal{I},\mathcal{Z})   | \psi(\mathcal{I},\mathcal{Z}) \big\rangle  
&\big \langle \psi(\mathcal{I},\mathcal{Z})   | \psi(\mathcal{Z},\mathcal{Z})  \big\rangle&\big\langle \psi(\mathcal{I},\mathcal{Z})   | \psi(\mathcal{Z},\mathcal{I})  \big\rangle\\
  \big\langle \psi(\mathcal{Z},\mathcal{I})   | \psi(\mathcal{I},\mathcal{I})   \big\rangle& \big\langle \psi(\mathcal{Z},\mathcal{I})   | \psi(\mathcal{I},\mathcal{Z})   \big\rangle
&\big\langle  \psi(\mathcal{Z},\mathcal{I})   | \psi(\mathcal{Z},\mathcal{Z})   \big\rangle&\big \langle \psi(\mathcal{Z},\mathcal{I})   | \psi(\mathcal{Z},\mathcal{I})   \big\rangle \\
   \big\langle \psi(\mathcal{Z},\mathcal{Z})   | \psi(\mathcal{I},\mathcal{I}) \big\rangle  &\big\langle  \psi(\mathcal{Z},\mathcal{Z})   | \psi(\mathcal{I},\mathcal{Z})   \big\rangle
& \big\langle \psi(\mathcal{Z},\mathcal{Z})   | \psi(\mathcal{Z},\mathcal{Z}) \big\rangle  & \big\langle \psi(\mathcal{Z},\mathcal{Z})   | \psi(\mathcal{Z},\mathcal{I})   \big\rangle  \\
\end{array} \right).
\end{align} 
\end{widetext}


\section{ The results} \label {results}

\subsection{ The quantum  $\mathbb{Z}_N$ phase}

\begin{figure}[ht]
\center{\epsfig{figure=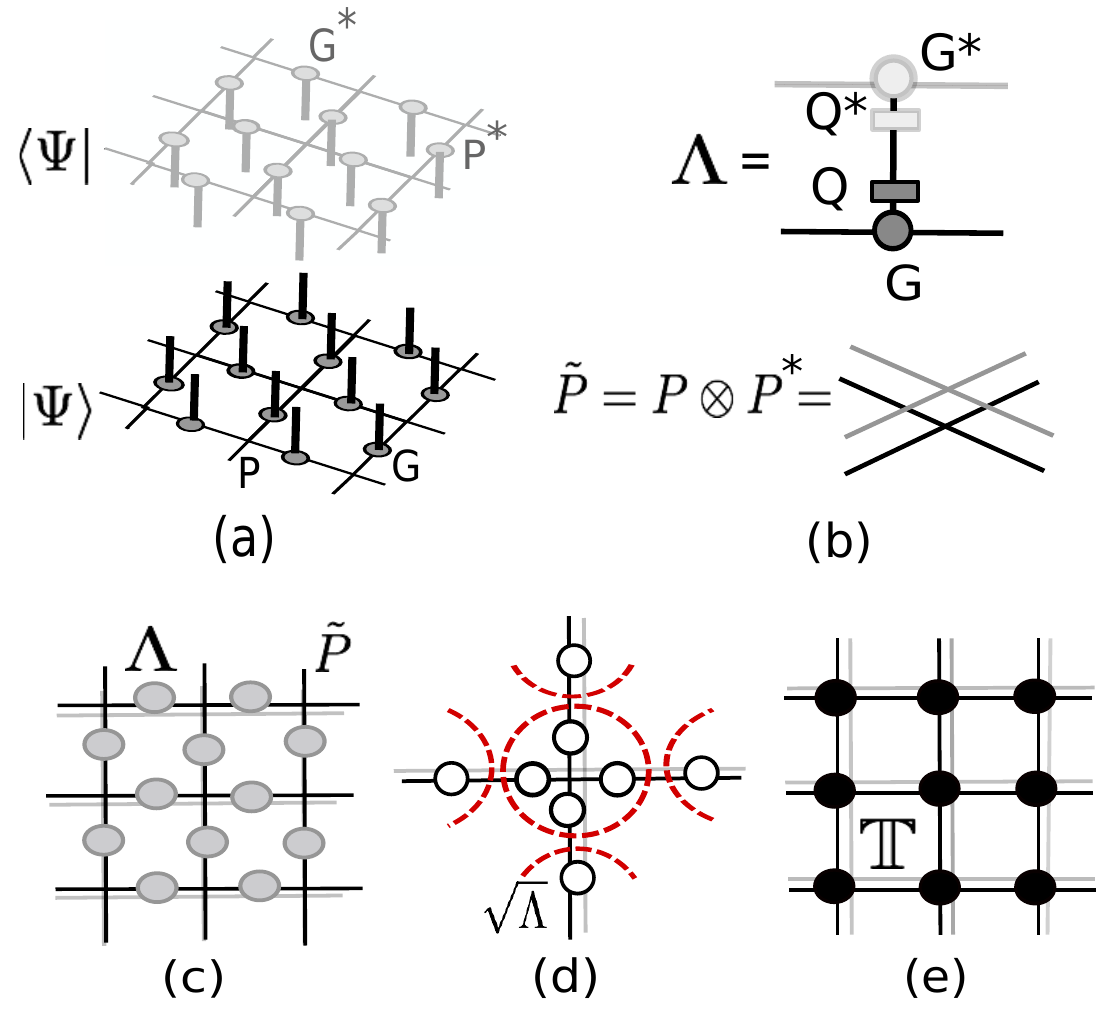,angle=0,width=8.5cm}}
\caption[]{(a) The tensor product state representation of $\mathbb{Z}_N$ topologically ordered phase, (b) Apply the deformation to form a double tensor $\Lambda$ on the link and the tensor on the vertex, $\tilde{P} = P \otimes P^*  $, (c) The double tensor represented by $\Lambda$ and $\tilde{P}$, (d)Decompose tensor $\Lambda= \sqrt{\Lambda}  \sqrt{\Lambda} $ and reconstruct the tensor (e) The new double tensor.} \label{TO}
\end{figure}

Let us begin by describing the construction of $\mathbb{Z}_N$ wavefunctions. 
Their Hamiltonian is generalized from toric code model \cite{Zn_2012,He_Wen_2014}.
The tensor product state (TPS) on the square lattice motivated by the  $\mathbb{Z}_N$ topologically ordered phase  is characterized by the rank-4 tensor, $P_{\alpha, \beta, \gamma,\delta}$ with four internal indices  running over $0,1,2,...N-1$ on vertex and the rank-3 tensor  $G^{s}_{\alpha,\beta}$ with one physical index $s$ running over the $N$ possible spin states ${0,1,....(N-1)}$ on the link as shown in Fig.~\ref{TO} (a).
 The wave function is then given by
\begin{align}
|\psi \rangle= \sum_{\{s^i\}}\text{ tTr}( \otimes_v P \otimes_l G^{s_i}) \mid s_1,s_2,....\rangle, 
\label{ZN_TPS}
\end{align} 
 where $v$ labels vertices and  $l$ links. Specifically, 
\begin{align}
 P_{\alpha, \beta,\gamma,\delta} = \left\{ 
  \begin{array}{l l}
    1, & \quad \text{if $\alpha+\beta+ \gamma+\delta=0$ mod N}, \\
    0, & \quad \text{otherwise}, 
  \end{array} \right.
  \label{TTlabel}
\end{align} 
and
\begin{align}
&G_{ii}^{i}=1, \quad i=0,1,2,3,...,N-1,   \notag \\
& \text{others=0}. 
\label{GGlabel}
\end{align}

\begin{figure*}[ht]
\center{\epsfig{figure=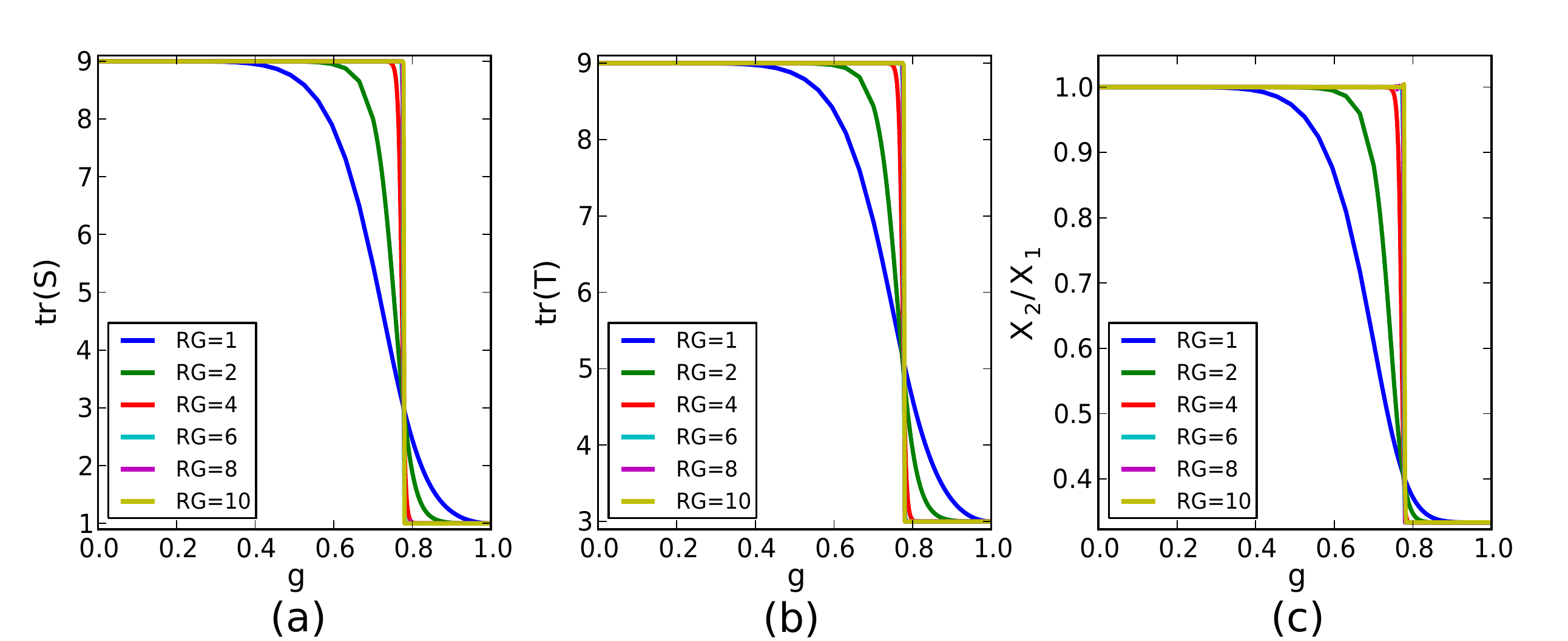,angle=0,width=17cm,height=6cm}}
\caption[]
{The trace of modular matrices  (a)$S$, (b) $T$, and the property  (c) $X_2/X_1$ as functions of parameter $g$ display a phase transition at critical point $g_c$  of  $\mathbb{Z}_3$   model}
   \label{Z3_phase}
\end{figure*}

The rank-3 tensor $G$ behaves like a projector which essentially sets the internal index equal to the physical index. 
In \cite{robustness_PEPS_cirac11}, they studied the problem of the stability of a tensor network state under physical perturbations to the local tensor. 
In view of this  we can consider a deformation, $Q=\sum_{i=0} ^{N-1} q_i |  i \rangle   \langle i |$ and $   0 \leq q_i \leq 1$ which apply to the physical indices,  $|\psi_{(Q)} \rangle  \equiv Q \otimes Q \otimes...  \otimes Q          | \psi \rangle $. 
At $q_i= 1, (i=0,1,2,...,N-1)$, this is exactly $\mathbb{Z}_N$ topologically ordered phases. 
At $q_0=1, \quad  q_i=0, (i=1,2,3,...,N-1)$, the tensor represent a product state of all $0$. 
At some critical point in parameters $q_i$, the phase transition will occur.  
Here, we first consider $q_0=1$ and $q_i=g^2, (i=1,2,3,...,N-1)$.

\begin{figure}[ht]
\center{\epsfig{figure=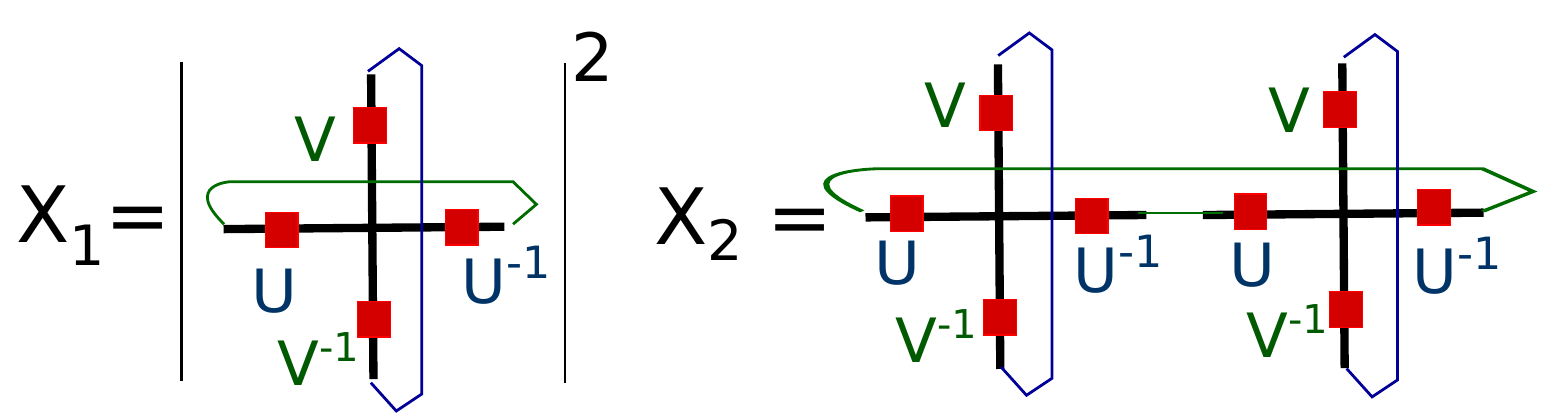,angle=0,width=7cm}}
\caption[]
{The quantity $X_2/X_1$ obtained by taking the ratio of the contraction value of the double tensor in two different ways. $X_2 / X_1$ is invariant under gauge transformation, such as unitary operators $U$ and $V$. It can be used to distinguish different fixed-point tensors. }
   \label{X2X1}
\end{figure}

 The $\mathbb{Z}_N$ phase has a $N^2$-fold ground-state degeneracy on a torus, which corresponds to $N^2$ different types of quasiparticle excitations. Therefore, the corresponding modular matrices will be of size $N^2 \times N^2$.
For simplicity of the calculation we associate every vertex with four matrices as shown in Fig.~\ref{TO} and   from the double tensor.
The norm of wave function represented by the double tensor can then be represented as standard tensor product  form as Eq.~(\ref{ST_TPS}), where the double tensor $\mathbb{T}^{\alpha', \beta', \gamma', \delta'}_{\alpha, \beta, \gamma, \delta}$ has eight inner indices $ \alpha', \beta', \gamma', \delta',\alpha, \beta, \gamma, \delta = 0,1,...,N-1 $. 
%
%
%
%
From the  TPS with deformation,  we can find the phase transition point of the $\mathbb{Z}_3$ model  as shown in Fig.~\ref{Z3_phase} by using GSPTRG. 
The fixed point tensor structure might be complicated but it is always possible to identify them. 
We calculate  $S$ and $T$ matrices  and a basis independent quantity given by the ratio $X_2/X_1$~\cite{Xie_LU_2010}, where 
$X_1$ and $X_2$ as shown in Fig. \ref{X2X1} are defined as follows,
\begin{align}
& X_1 =    \big(   \sum _{s,\alpha, \beta, \alpha', \beta'} A^s_{\alpha, \beta, \alpha, \beta}   \times  (A^{s}_{\alpha', \beta', \alpha', \beta'} )^*      \big) ^2, \notag \\
& X_2 =  \sum _{s,s',\alpha, \beta,\gamma,\delta, \alpha', \beta',\gamma',\delta'} 
(A^s_{\alpha, \beta, \gamma, \beta} \times A^{s'}_{\gamma, \delta,  \alpha, \delta})
\times \notag \\
&\big( (A^s_{\alpha', \beta', \gamma', \beta'})^*   \times  \big( A^{s'}_{\gamma', \delta',  \alpha', \delta'})^*   \big). 
\end{align}

We find that when $0 \leq g < 0.7776 $, all components of $S$ and $T$ matrices are $1$, and $X_2/X_1=1.0$, and this shows that the ground state is in the trivial  phase. 
When  $0.7776 \leq g < 1.0 $,  the tensor belongs to the $\mathbb{Z}_3$ topologically ordered phase, since we obtain nontrivial $S$ and $T$ matrices as follows:
\begin{align} 
\label{Z3_Smatrix}
S =  \left(
\begin{array}{cccccccccc}
  1 & 0  &  0 & 0 & 0  & 0  & 0 & 0 &0  \\
  0 & 0  &  0 & 0 & 0  & 0  & 1 & 0 &0  \\
  0 & 0  &  0 & 1 & 0  & 0  & 0 & 0 &0  \\
  0 & 1  &  0 & 0 & 0  & 0  & 0 & 0 &0  \\  
  0 & 0  &  0 & 0 & 0  & 0  & 0 & 1 &0  \\
  0 & 0  &  0 & 0 & 1  & 0  & 0 & 0 &0  \\
  0 & 0  &  1 & 0 & 0  & 0  & 0 & 0 &0  \\
  0 & 0  &  0 & 0 & 0  & 0  & 0 & 0 &1  \\
  0 & 0  &  0 & 0 & 0  & 1  & 0 & 0 &0  \\
\end{array} \right),
\end{align} 
\begin{align} 
\label{Z3_Tmatrix}
T =  \left(
\begin{array}{cccccccccc}
  1 & 0  &  0 & 0 & 0  & 0  & 0 & 0 &0  \\
  0 & 1  &  0 & 0 & 0  & 0  & 0 & 0 &0  \\
  0 & 0  &  1 & 0 & 0  & 0  & 0 & 0 &0  \\
  0 & 0  &  0 & 0 & 0  & 1  & 0 & 0 &0  \\  
  0 & 0  &  0 & 1 & 0  & 0  & 0 & 0 &0  \\
  0 & 0  &  0 & 0 & 1  & 0  & 0 & 0 &0  \\
  0 & 0  &  0 & 0 & 0  & 0  & 0 & 1 &0  \\
  0 & 0  &  0 & 0 & 0  & 0  & 0 & 0 &1  \\
  0 & 0  &  0 & 0 & 0  & 0  & 1 & 0 &0  \\
\end{array} \right), 
\end{align} 
and  $X_2/X_1=0.33333$.
The $S$ and $T$ matrices obtained above give us the modular transformations.
Note that the matrices are not in the canonical form (where, e.g., the $T$ is diagonal),  but
there  is a procedure to make the $T$-matrix diagonal and  at the same time make $S$ in the canonical form~\cite{Liu_2013}.
Then diagonalized  $T$-matrix gives the self-statistics of quasiparticles, and $S$ matrix the mutual statistics.

In the GSPTRG algorithm, we preserve the  $\mathbb{Z}_3$ gauge symmetry of the tenser. 
As performing more steps of RG, the nonlocal order parameters show sharper changes around $g_c=0.7776$ in Fig.~\ref{Z3_phase}, and 
the crossings of different RG curves signal the transition point separating the trivial phase and topological phase.

\begin{figure}[ht]
\center{\epsfig{figure=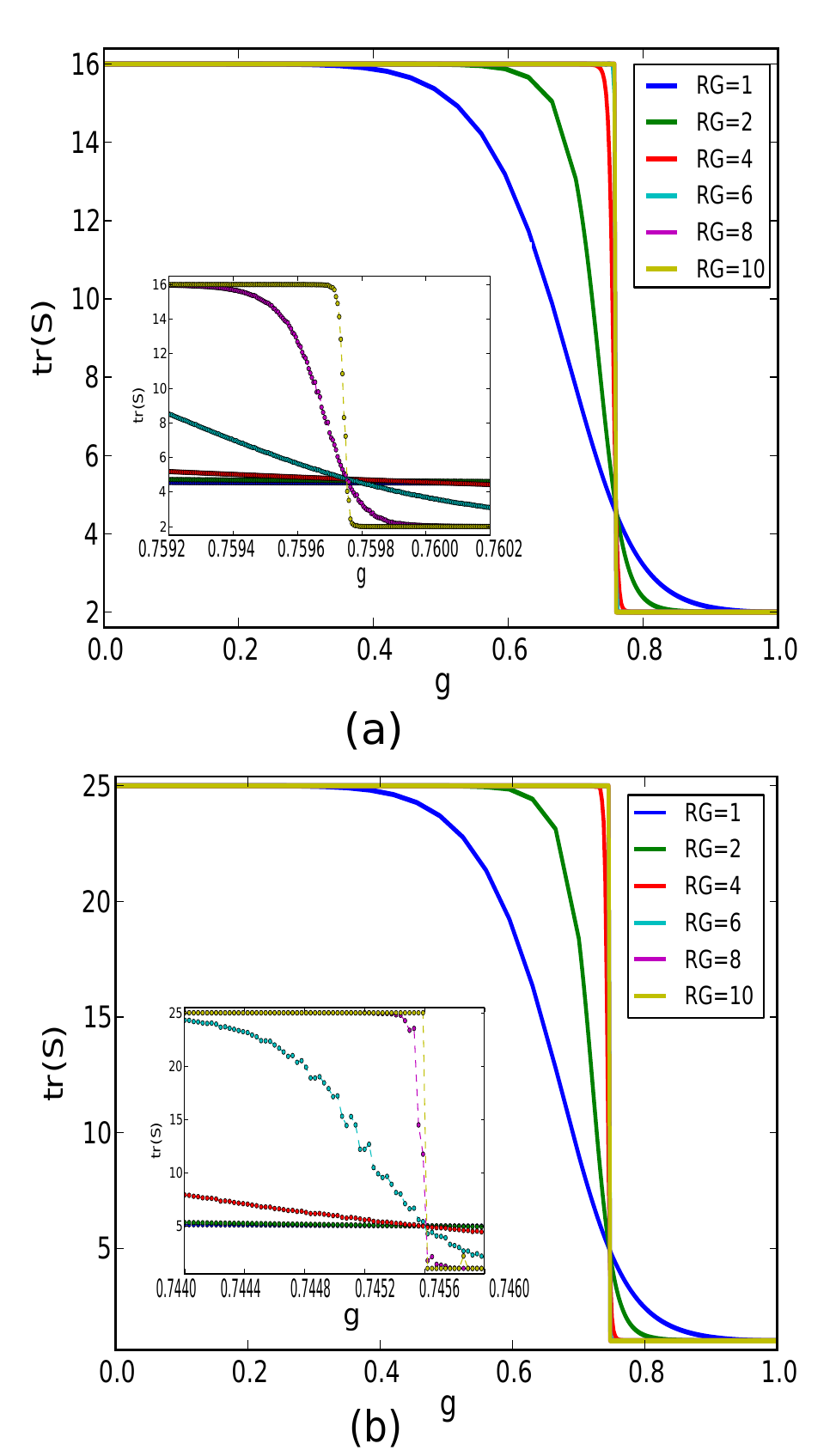,angle=0,width=8.5cm,height=11cm}}
\caption[]{The trace of modular matrices  $S$ as functions of parameter $g$ display a phase transition at critical point $g_c$  of   (a)$\mathbb{Z}_4$  (b)  $\mathbb{Z}_5$  model  } \label{ZN_total_S}
\end{figure}

Similar behaviors of the  $\mathbb{Z}_4$ and   $\mathbb{Z}_5$ model under GSPTRG is also found  in Fig.~\ref{ZN_total_S}.
For $\mathbb{Z}_4$ model, when  $0.7597 \leq g < 1.0 $,  the tensor belongs the $\mathbb{Z}_4$ topologically ordered phase, since we obtain nontrivial $S$ and $T$ matrices as follows:
\begin{align}  
S =
\begin{pmatrix}
1&&&&&&&&&&&&&&&\\
&&&&1&&&&&&&&&&&\\
&&&&&&&&&&&&1&&&\\
&&&&&&&&1&&&&&&&\\
&1&&&&&&&&&&&&&&\\
&&&&&&&&&&&&&&&1\\
&&&&&&&&&&&1&&&&\\
&&&&&&&1&&&&&&&&\\
&&1&&&&&&&&&&&&&\\
&&&&&&&&&&1&&&&&\\
&&&&&&&&&&&&&1&&\\
&&&&&1&&&&&&&&&&\\
&&&1&&&&&&&&&&&&\\
&&&&&&&&&&&&&&1&\\
&&&&&&&&&1&&&&&&\\
&&&&&&1&&&&&&&&&\\
\end{pmatrix},
\end{align} 
\begin{align}  
T =
\begin{pmatrix}
1&&&&&&&&&&&&&&&\\
&1&&&&&&&&&&&&&&\\
&&1&&&&&&&&&&&&&\\
&&&1&&&&&&&&&&&&\\
&&&&&&&1&&&&&&&&\\
&&&&&&1&&&&&&&&&\\
&&&&&1&&&&&&&&&&\\
&&&&1&&&&&&&&&&&\\
&&&&&&&&&1&&&&&&\\
&&&&&&&&&&&1&&&&\\
&&&&&&&&1&&&&&&&\\
&&&&&&&&&&&&&&&\\
&&&&&&&&&&&&&1&&\\
&&&&&&&&&&&&&&&1\\
&&&&&&&&&&&&1&&&\\
&&&&&&&&&&&&&&1&\\
\end{pmatrix},
\end{align} 
and  $X_2/X_1=0.25$ as shown in Fig.~\ref{ZN_total_S} (a).
For $\mathbb{Z}_5$ model, when  $0.7450 \leq g < 1.0 $,  the tensor belongs the $\mathbb{Z}_5$ topologically ordered phase, since we obtain nontrivial $25 \times 25$ $S$ and $T$ matrices and  $X_2/X_1=0.2$ as shown in Fig.~\ref{ZN_total_S} (b).

In the following, we shall find the mapping from the norm square of the $\mathbb{Z}_N$ wavefunctions and the $N$-state Potts model. 
First, by applying a deformation $Q=\sum_{i=0} ^{N-1} q_i |  i \rangle   \langle i |$  to tensor  $G^{s}_{\alpha,\beta}$,  a new tensor $\Lambda $  can be obtained as shown in Fig.~\ref{TO} (b): 
\begin{align}
\Lambda_{\alpha,\beta}^{\alpha', \beta'} = \sum_{s,s',s''}   Q_{s,s'}  G^{s'}_{\alpha \beta} \times Q^*_{s, s''}  G^{*s''}_{\alpha,\beta}.
\end{align} 
There are two nonzero components in $\Lambda_{\alpha,\alpha'}^{\beta,\beta'}$,
\begin{align}
\Lambda_{0,0}^{0,0} = 1;   \quad   \Lambda_{i, i}^{i, i} = q_i^2   \quad (i=1,2,...,N-1).
\end{align}

Second, we form the double tensor $\mathbb{T}$ as shown in Fig.~\ref{TO} (e)  from tensors $\tilde{P} = P \otimes P^{*}  $ on the vertex and $\Lambda$ on the link as shown in Fig.~\ref{TO} (d),  
\begin{align}
\mathbb{T}_{i,j, k,l}^{i', j', k', l'} = & \sum _{ \alpha, \beta, \gamma,\delta, \alpha ', \beta ', \gamma ',\delta '}    P_{\alpha, \beta, \gamma, \delta} \times P^*_{\alpha ',  \beta ',  \gamma ', \delta '}   \notag \\ 
 &\sqrt{ \Lambda_{\alpha,i} ^{\alpha', i '}} 
 \sqrt{ \Lambda_{\beta, j} ^{\beta', j '}}  
 \sqrt{ \Lambda_{\gamma, k}^{\gamma ',  k '}}  
 \sqrt{ \Lambda_{\delta,l} ^{\delta', l '}}, 
 \end{align} 
 \begin{align}
\mathbb{T}_{i,j,k,l}^{i,j,k,l}  = \prod_{m=0}^{N-1}  q_m^{ n_m},
\end{align} 
where $n_m$ is the number of virtual indices in state "$m$".
Then the double tensor with nonzero components of  $\mathbb{Z}_2$ model for $Q=|  0 \rangle   \langle 0 |  + g^2 | 1 \rangle   \langle 1 | $ are given by
\begin{align}
\label{Z2quantum}
& \mathbb{T}_{0000}^{0000} =  1  \\
&  \mathbb{T}_{0011}^{0011} = \mathbb{T}_{0110}^{0110}= \mathbb{T}_{1100}^{1100}   \notag \\
& =\mathbb{T}_{1001}^{1001} = \mathbb{T}_{0101}^{0101} = \mathbb{T}_{1010}^{1010} =  g^4     \notag \\
&  \mathbb{T}_{1111}^{1111} =  g^8   \notag. 
\end{align} 
The double tensor with nonzero components  of  $\mathbb{Z}_3$ model for   $Q=|  0 \rangle   \langle 0 |  + g^2 | 1 \rangle   \langle 1 | + g^2 | 2 \rangle   \langle 2 | $ are given by
\begin{align}
\label{Z3quantum}
&  \mathbb{T}_{0000}^{0000} =  1  \\
&  \mathbb{T}_{1110}^{1110}=  \mathbb{T}_{1101}^{1101}=  \mathbb{T}_{1011}^{1011}=  \mathbb{T}_{0111}^{0111}=  g^6   \notag \\
& \mathbb{T}_{2220}^{2220}= \mathbb{T}_{2202}^{2202}= \mathbb{T}_{2022}^{2022}= \mathbb{T}_{0222}^{0222} =  g^6     \notag \\
& \mathbb{T}_{0012}^{0012} =   \mathbb{T}_{0120}^{0120} =   \mathbb{T}_{1200}^{1200} =  \mathbb{T}_{2001}^{2001} =  \mathbb{T}_{1002}^{1002} =   \mathbb{T}_{0021}^{0021}      \notag \\
& =\mathbb{T}_{0210}^{0210} =  \mathbb{T}_{2100}^{2100} =  \mathbb{T}_{0102}^{0102} =  \mathbb{T}_{1020}^{1020} =  \mathbb{T}_{0201}^{0201} =   \mathbb{T}_{2010}^{2010} =  g^4     \notag \\
&\mathbb{T}_{1122}^{1122} =  \mathbb{T}_{1221}^{1221} =  \mathbb{T}_{2211}^{2211} = \mathbb{T}_{2112}^{2112} =  \mathbb{T}_{1212}^{1212} =  \mathbb{T}_{2121}^{2121} =  g^8  \notag .
\end{align}

  The $\mathbb{Z}_2$ model above is mathematically equivalent to the two-dimensional classical Ising model where the transition point is known to great accuracy~\cite{Xie_LU_2010}.
The numerical results from Ref.~\cite{He_Wen_2014}  show that the $\mathbb{Z}_2$ toric code and double-semion models have the same critical point under the string tension. 
This is understood by recognizing that the phases in the double-semion wavefunction  of Ref.~\cite{He_Wen_2014} under the string tension cancel in the mapping to the Potts model and its has the same norm square as the one from deforming the toric code. Therefore, they have the same critical point $g_c=0.802243$ from Eq.~(\ref{eqn:gc}) below.  
Next we derive the mapping.

 \subsection{The classical $N$-state Potts Model}
 
The Hamiltonian of the $N$-state Potts is given by 
\begin{align}
H = \sum _{<i,j>} (-J \delta_{s_i,s_j} ).
\end{align} 
with the sum running over the nearest neighbor pairs $<i, j>$ over all lattice sites. 
The degree of freedom $s_i,s_j$ on the site is on values in $\{0,1,..., q-1\}$. 
The $\delta_{s_i,s_j}$  is the Kronecker delta, which equals one whenever $s_i = s_j$ and zero otherwise. 
The model is ferromagnetic when  $J>0$ and antiferromagnetic if $J<0$.
A basic question is where the phase transition point is. 
Baxter has determined the exact free energy for the square-lattice Potts model and determined the critical point, such as ferromagnetic critical point is $ e^{\beta J}=1+\sqrt{q}$ and antiferromagnetic critical point is $ e^{\beta J}=-1+\sqrt{4-q}$~\cite{Baxter88}.

For example, the three-state Potts model with zero external field is  $N=3$ and then the spins are usually taken to be $\{0,1,2\}$. 
By tuning the temperature, the phase transition will occur, 
for example, for $J > 0$ the transition is first order  if $N \geq 5$ and is  continuous  if $N \leq 4$.

The partition function of  $N$-state Potts Model is given by:
\begin{align}
Z = e ^{-\beta H} = \sum \prod _{<i,j>} e ^{\beta J  \delta_{s_i,s_j} } = \sum \prod_{<i,j>} \Lambda(i,j).  
\end{align} 
Here, $\Lambda$ is named the transfer Matrix on the link and can be represented by a $N \times N$ matrix

\begin{align}  
\Lambda (i,j)=
\begin{pmatrix}
    e^{\beta J} & 1 & 1 & \dots  & 1 \\
    1 & e^{\beta J}  & 1 & \dots  & 1 \\
    \vdots & \vdots & \vdots & \ddots & \vdots \\
    1 & 1 & 1 & \dots  & e^{\beta J} 
\end{pmatrix}. 
\end{align}

\begin{figure}[ht]
\center{\epsfig{figure=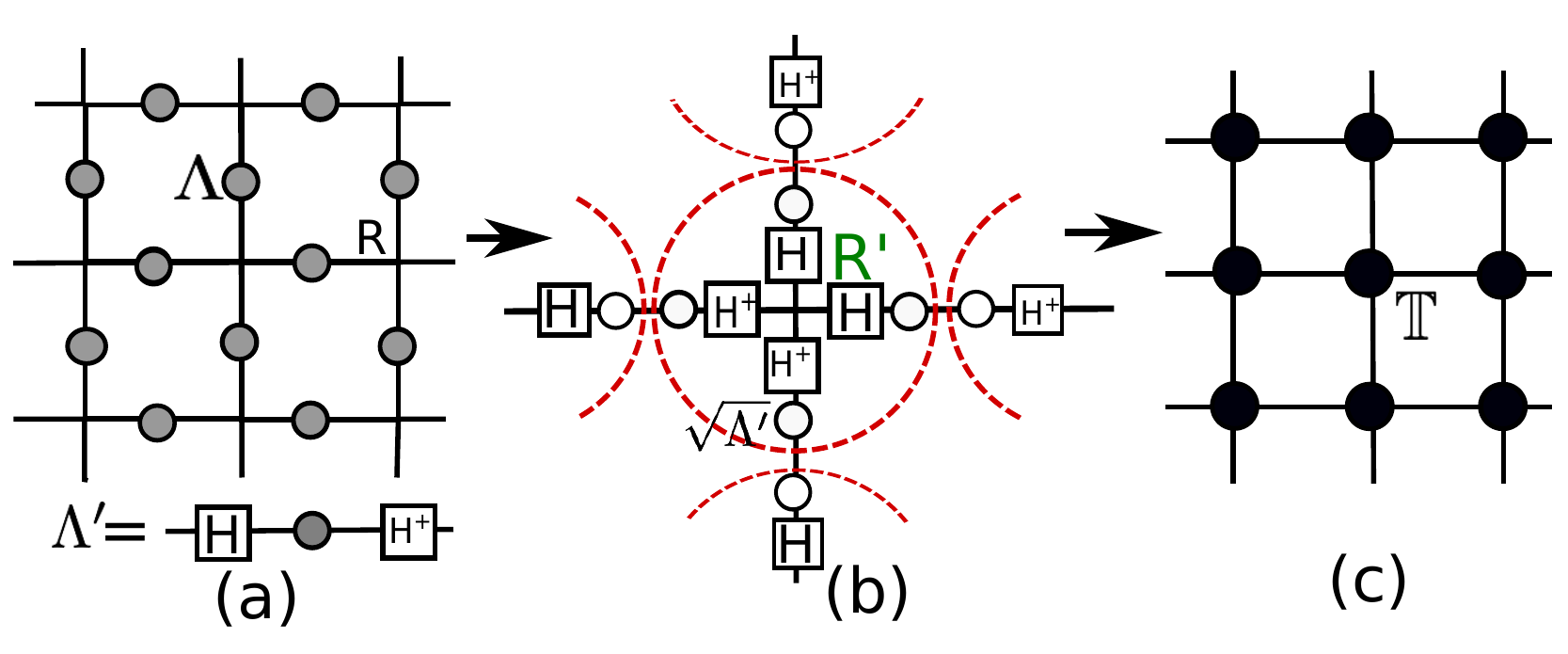,angle=0,width=8.5cm}}
\caption[] {The tensor representation of partition function, (b) Reconstruct the tensor (c) The new tensor configuration.} \label{classical}
\end{figure}

Applying the Hadamard matrix $H$, we form a diagonal matrix $\Lambda'$ as shown in Fig.~\ref{classical} (a). 
\begin{align} 
\Lambda' =& H\times \Lambda  \times H^{\dagger}  \notag\\
= &
\begin{pmatrix}
    e^{\beta J}+q-1 & 0 & 0 & \dots  & 0 \\
    0 & e^{\beta J} -1 & 0 & \dots  & 0 \\
    \vdots & \vdots & \vdots & \ddots & \vdots \\
    0 & 0 & 0 & \dots  & e^{\beta J}-1 
\end{pmatrix}. 
\end{align} 
For example, the Hadamard matrix of three-state Potts model can be given by 
\begin{align} 
H = \sum_{\alpha,\beta =0 }^2 H_{\alpha \beta} = \frac{1}{\sqrt3}  \left(
\begin{array}{cccc}
  1 & 1   &  1  \\
  1& \omega  & \omega^2 \\
  1&  \omega^2 & \omega  \\
\end{array} \right),
\end{align} 
where $\omega = e^{i 2 \pi /  3}$. 

Then, we define a trivial rank-4 tensor $R_{\alpha  \alpha  \alpha \alpha}=1$ with index $\alpha$  running over $0,1,2,...,N-1$ on each vertex. 
To apply the Hadamard matrix $H$ to the tensor $R$, 
\begin{align}
R'_{i,j,k,l} 
&=  \sum _{ \alpha} R_{\alpha,  \alpha, \alpha,\alpha} \times  H_{\alpha, i} H^{+}_{\alpha, j} H^{+}_{\alpha, k}H_{\alpha,l} \\
& = \left\{ 
  \begin{array}{l l}
    1/q, & \quad \text{if $\alpha+\beta+ \gamma+\delta=0$ mod N}\\
    0, & \quad \text{otherwise}. 
  \end{array} \right.  
\end{align} 
Combining the tensor $R'$ and $\Lambda '$  forms a new tensor $\mathbb{T}$. 
Then the double  tensor with nonzero components of classical Ising model without magnetic field are
\begin{align}
\label{Z2classical}
& \mathbb{T}_{0000} =  \frac{1}{2} (2\cosh(\beta J))^2  \notag \\
& \mathbb{T}_{0011} = \mathbb{T}_{0110} = \mathbb{T}_{1100} = \frac{1}{2} (2\cosh(\beta J))(2\sinh(\beta J))     \notag \\
&\mathbb{T}_{1001} =\mathbb{T}_{0101} =\mathbb{T}_{1010} = \frac{1}{2} (2\cosh(\beta J))(2\sinh(\beta J))     \notag \\
& \mathbb{T}_{1111} =  \frac{1}{2}  (2\sinh(\beta J))^2.
\end{align}

The nonzero components of the three-state Potts model are given by 
\begin{align}
\label{Z3classical}
& \mathbb{T}_{0000} =  \frac{1}{3} (\sqrt{e^{\beta J}+2 } )^4  \notag \\
& \mathbb{T}_{0111} =  \mathbb{T}_{1011} =  \mathbb{T}_{1101} =  \mathbb{T}_{1110} 
    =   \frac{1}{3}  (\sqrt{e^{\beta J}+2 } )  (\sqrt{e^{\beta J}-1 } )^3  \notag \\   
& \mathbb{T}_{0222} =   \mathbb{T}_{2022} =  \mathbb{T}_{2202} =  \mathbb{T}_{2220} 
     =  \frac{1}{3} (\sqrt{e^{\beta J}+2 } )  (\sqrt{e^{\beta J}-1 } )^3 \notag \\
& \mathbb{T}_{0012}  = \mathbb{T}_{0120}  = \mathbb{T}_{1200}  = \mathbb{T}_{2001} =   \notag \\ 
& \mathbb{T}_{0021} =  \mathbb{T}_{0210}=   \mathbb{T}_{2100} =  \mathbb{T}_{1002} =   \notag \\ 
&  \mathbb{T}_{0102} = \mathbb{T}_{1020} = \mathbb{T}_{0201} = \mathbb{T}_{0201} =
     \frac{1}{3} (\sqrt{e^{\beta J}+2 } )^2  (\sqrt{e^{\beta J}-1 } )^2 \notag \\ 
& \mathbb{T}_{1122} =   \mathbb{T}_{1221} =   \mathbb{T}_{2211} =   \mathbb{T}_{2112} =  \notag \\ 
&    \mathbb{T}_{1212} =   \mathbb{T}_{2121} =  
    \frac{1}{3} (\sqrt{e^{\beta J}-1 } )^4. 
\end{align}

\begin{figure*}[ht]
\center{\epsfig{figure=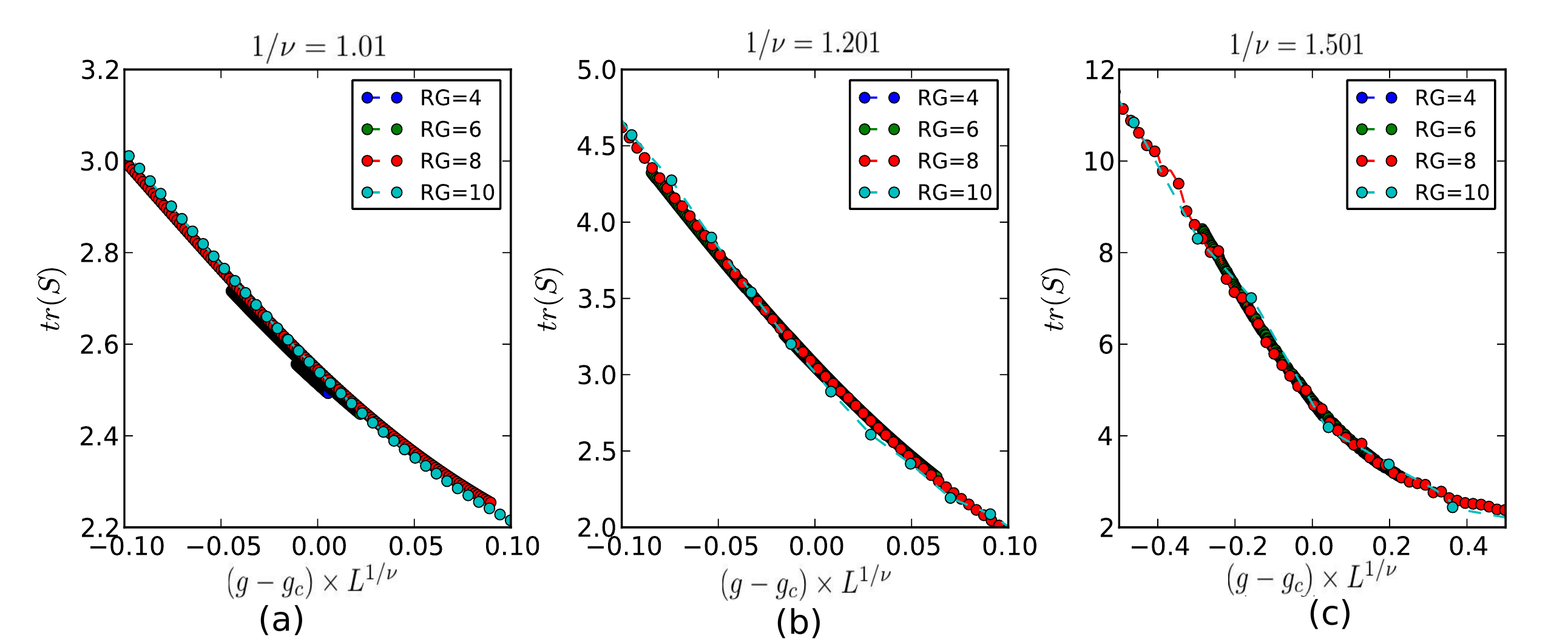,angle=0,width=17cm,height=6cm}}
\caption[]
{  The trace of $S$ matrix as a function $(g-gc)\times L^{\nu}$ on  (a)$\mathbb{Z}_2$  (b)$\mathbb{Z}_3$  (c)  $\mathbb{Z}_4$  models.   }
   \label{CP_total}
\end{figure*}

\subsection {Duality}

The two-dimensional classical ferromagnetic Potts models have phase transitions 
%
located at  $e^{(\beta_c J)} =1+ \sqrt{q}$,  separateing the ordered (ferromagnetic) and disordered (paramagnetic) phases.  
The form of   partition function of the $N$-state Potts model  represented by the tensor network is equal to the norm square of the  $\mathbb{Z}_N$ wave function, with deformation $ Q = | 0 \rangle   \langle 0 |   + \sum_{i=1}^{N-1}g^2 | i \rangle   \langle i |$, represented by the tensor network, such as Eq.~(\ref{Z2quantum}), Eq.~(\ref{Z2classical}) and  Eq.~(\ref{Z3quantum}), Eq.~(\ref{Z3classical}).
The double tensor of the norm of  $\mathbb{Z}_N$ wave function is just a two copies of the partition function of Potts model. 
We then have the relation between the parameter $g$ and the parameter $\beta J$,
\begin{align}
\label{eqn:gc}
g  = \left(  \frac{  \sqrt{e^{\beta J}-1 }  ^2 } {\sqrt{e^{\beta J}+N-1}  ^2 } \right) ^{1/8}.
\end{align} 

From the relation and the  transition point of Potts model, we can then obtain the phase transition point $g_c$ for $\mathbb{Z}_N$ model :
\begin{center}
\begin{tabular}{l  |c  | c ||  c | c }  
             & Numerics   & \scriptsize From mapping   & Numerics  &    \scriptsize  Potts model   \\
\hline
N             & $ g_c $   & $g_c(\beta J)$&  $ 1/ \nu$ & $ 1/ \nu$    \\
\hline
\hline 
2 & 0.8021  \tiny($\chi_c=24$) &  0.802243 &  1.010   &   1         \\
3 & 0.7776  \tiny($\chi_c=24$) &  0.777817 &  1.201   &   6/5       \\
4 & 0.7597  \tiny($\chi_c=32$) &  0.759835  &   1.501 &   3/2    \\
5 & 0.7450   \tiny($\chi_c=30$)  &  0.745582 &     &    first order
 \label {CP_val}
\end{tabular}
\end{center}
From this table, the transition points from the GSPTRG are quite  close to exact mapping results.

For the classical $N$-state Potts Model, if $N \leq 4$, this model describes a continuous phase transition with scalar order parameter. 
The critical exponents of these transition are universal values and characterize the singular properties of physical quantities. 
In particular,  the correlation function of the classical spin system is, 
\begin{align}
D(\vec{r_i} - \vec{r_j}) = \langle  S(\vec{r_i} ) S(\vec{r_j} ) \rangle -  \langle S(\vec{r_i} )  \rangle  \langle S(\vec{r_j} )  \rangle, 
\end{align} 
where the brackets mean statistical average over all configurations.  
The correlation length $\xi$ is defined in terms of correlation function $D(\vec{r_i} -\vec{r_j} ) \sim  e^{|\vec{r_i} -\vec{r_j} |/ \xi }$, where  $|\vec{r_i} -\vec{r_j}|$ is the distance between two spins. 
In the asymptotic limit of large  $|\vec{r_i} -\vec{r_j}|$, the correlation function decays to zero exponentially.
On the other hand, the correlation length diverges at the critical point $\xi  \sim |T-T_c| ^{-\nu}$,  where $\nu$ is an example of a critical exponent. 
In the above Table, we list the exponents  for the classical  Potts Model.

As shown above the $\mathbb{Z}_N$ model with one parameter $g$ can be mapped to the classical $N$-state Potts Model. 
We fit critical exponent $\nu$ for the order parameter i.e. $Tr(T), Tr(S),$ or $X_2/X_1$ under the renormalization flow. 
Our results as shown in Fig.~\ref{CP_total} are $1/\nu=1.01$ for the $\mathbb{Z}_2$ model,  $1/\nu=1.201$ for the $\mathbb{Z}_3$ model and $1/\nu=1.501$ for the $\mathbb{Z}_4$ model. 
For the five-state Potts model, the transition is a weak first-order~\cite{Potts_1982}. 
 Near critical point, the correlation length is very large but finite.  
It is difficult to distinguish the critical exponent of  the $\mathbb{Z}_5$ model numerically. 
It is worth mentioning that since after each step of the renormalization, the number of sites is reduced by half, this is to say that,  after performing two steps of renormalization, the distance between two new neighbor sites will be increased twofold.
We can perform a rescaling, and see the data collapse. 
In Fig.~\ref{CP_total}, $L$ is length scale defined by the number of RG step  $n_{rg}$,  $L = 2^{n_{rg}/2}$.

\section {The critical phase} \label{critical}

Here we focus on the $\mathbb{Z}_3$ topologically ordered phase under different forms of deformation. 
As mentioned above,  the $\mathbb{Z}_3$ topologically ordered phase can be represented by rank-four tensor $P$ (see Eq.~(\ref{TTlabel}) ) and rank-three tenors $G$ (see  Eq.~(\ref{GGlabel}) ). 
We have already discussed a special projector $Q_a= 1  \times |0 \rangle   \langle 0  | + g^2 \times |1 \rangle   \langle 1  |  +g^2 \times  |2 \rangle   \langle 2 |   $ with $0 \leq g \leq 1$ in earlier parts, and here only briefly review it.
At some critical point in $g$, the state must go through a phase transition. 
Moreover,  this model is mathematically equivalent to two-dimensional three-state Potts model.
We can obtain  $g_c = 0.777817$ from exactly mapping. 
By using GSPTRG, we can obtain the transition point to be between $0.7776$ and $0.7777$ and find our results to be with $0.02\% $ accuracy.   

We shall discuss the more deformation operator $Q=  q_0\times  |0 \rangle   \langle 0  | + q_1 \times |1 \rangle   \langle 1  |  + q_2 \times |2 \rangle   \langle 2 |   $ with $q_0, q_1, q_2 >0$  in detail. 
By applying it to the $\mathbb{Z}_3$ toric code wavefunction, we obtain the double tensor
\begin{align}
 \mathbb{T}_{i,j,k,l} ^{i,j,k,l} = \left\{ 
  \begin{array}{l l}
    q_0^{n_0 }  q_1^{n_1 }  q_2^{n_2 }, & \quad \text{if $i+j+k+l=0$ mod 3}\\
    0, & \quad \text{otherwise},
  \end{array} \right.
\end{align}
where $i,j,k,l=0,1,2$ and $n_0$, $n_1$, and $n_2$ means the number of the inner indices in state "$0$", "$1$", and "$2$" respectively.


\begin{figure*}[ht]
\center{\epsfig{figure=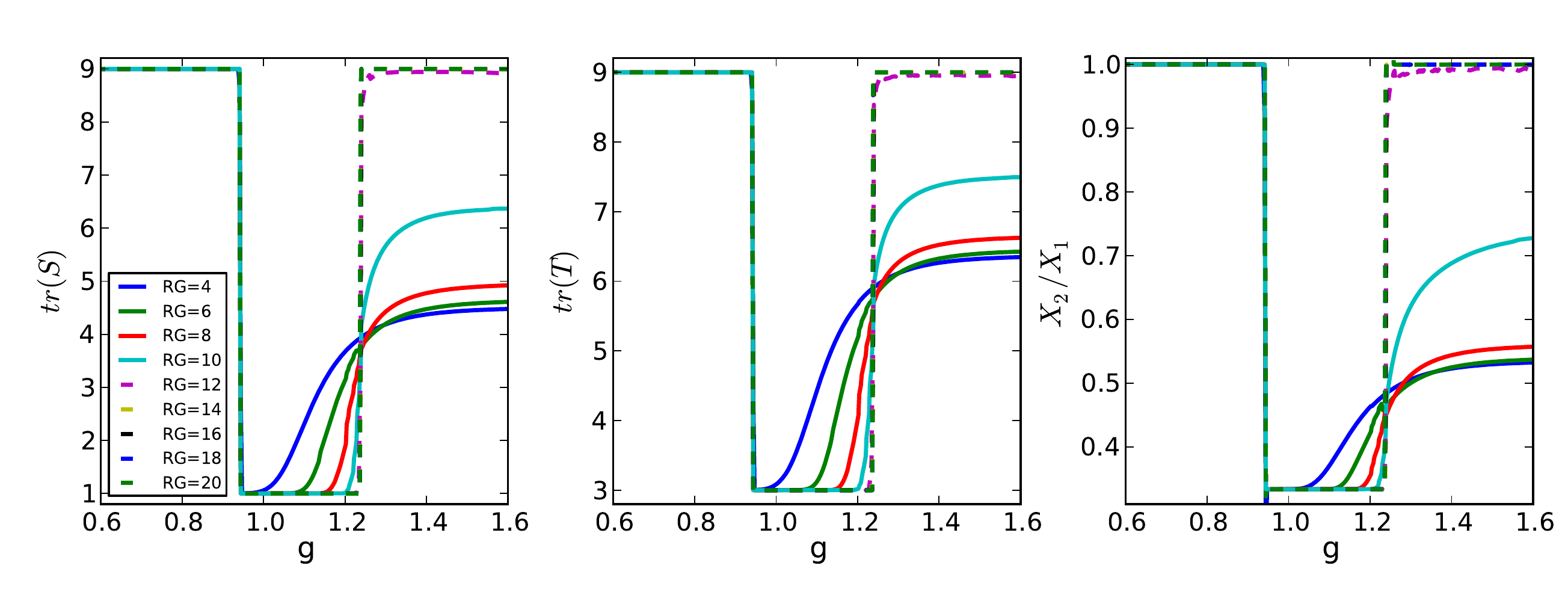,angle=0,width=17cm,height=6cm}}
\caption[]
{The trace of modular matrices  (a)$S$, (b) $T$, and the property  (c) $X_2/X_1$ as functions of parameter $g$ display a phase transition at critical point $g_c$  of  $\mathbb{Z}_3$   model with projector  $Q_b= 1 \times  |0 \rangle   \langle 0  |  +0 \times | 1  \rangle   \langle 1 |  +g^2 \times |2 \rangle   \langle 2 | $. }
   \label{10g_phases_giagram}
\end{figure*}


\subsection {For $ q_0=1; q_1=0; q_2=g^2$ }

We first study the case $ q_0=1; q_1=0; q_2=g^2$, i.e., the wavefunction is constructed from an effective two-dimensional Hilbert space spanned by $|0\rangle$ and $|2\rangle$. 
At $g=0$, the tensor represents a product of all states being $0$. Therefore, the phase diagram near $g=0$ is a region of the trivial phase that is adiabatically connected to a product state.
At $g> 0$, the nonzero components are 
\begin{align}
 &\mathbb{T}_{0222} ^{0222} = g^6,   \mathbb{T}_{2022} ^{2022} = g^6;  \mathbb{T}_{2202} ^{2202} = g^6 ; \mathbb{T}_{2220} ^{2220} = g^6 \notag \\
 & \mathbb{T}_{0000} ^{0000} = 1. 
\end{align} 
While $g \gg 1$, the tensor form is mathematically equivalent the quantum dimer model at Rokhsar-Kivelson (RK) point \cite{RK1988,Moessner2001,Summary_QDM} where the degenerate ground state attains exactly zero energy in the form of an equal weight superposition of all possible configurations in a given winding parity sector on square lattice.
This point is a critical phase with algebraically  decaying correlation function.  
It is noteworthy that  the RK point is a critical point separating two gapped phases of the quantum dimer model on the square lattice.
In the large $g$ limit, we can regard string-$2$ as vacuum state and string-$0$ as dimer. Then the tensor form will  obey the hard-core constraint where each vertex is connected to one dimer only and has the same weight.

As a result, as $ g$ goes from $0$ to $2$, a phase transition must occur. 
Some phases between gapped trivial states and critical phases could exist. Can a topologically ordered phase exist?
Let us examine this middle parameter region in detail.

We use the gauge symmetry preserved tensor renormalization group to flow our wave function to fixed point and  obtain the modular matrices by inserting gauge transformation and invariant quantity $X_2/X_1$ easily with sharp quantum phases transition point. 
Again, this is a advantage to use GSPTRG to characterize the topologically ordered phase numerically with TPS ansatz.
Our numerical results is given in Fig.~\ref{10g_phases_giagram}.

We see that when  $0 \leq g < 0.944$, all components of  $9\times 9$ $S$ and $T$ matrices are $1$, and $X_2/X_1 = 1.0$.  We can say the ground state in the trivial (product) phase, since  at $g=0$ this wave function is product state.
For $0.944 \leq  g < 1.238$, the tensor belongs the $\mathbb{Z}_3$ topologically ordered phase, since we obtain nontrivial $S$ and $T$ matrices  (same as Eq.~(\ref{Z3_Smatrix}) and (\ref{Z3_Tmatrix})) and $X_2/X_1=0.3333$.
For $  1.238 \leq  g <2.0$, when renormalization step is larger enough, we can obtain the trivial fixed  point. 
The GSPTRG can distinguish different topologically ordered phases and a topologically ordered phase from topologically trivial phases. But it cannot distinguish different topologically trivial phases.  From the equivalence to the RK point for large $g$, the system is probably gapless for $g$ large; from the trivial product state at $g=0$, the system is probably also a trivial product phase for small $g$.

\begin{figure}[ht]
\center{\epsfig{figure=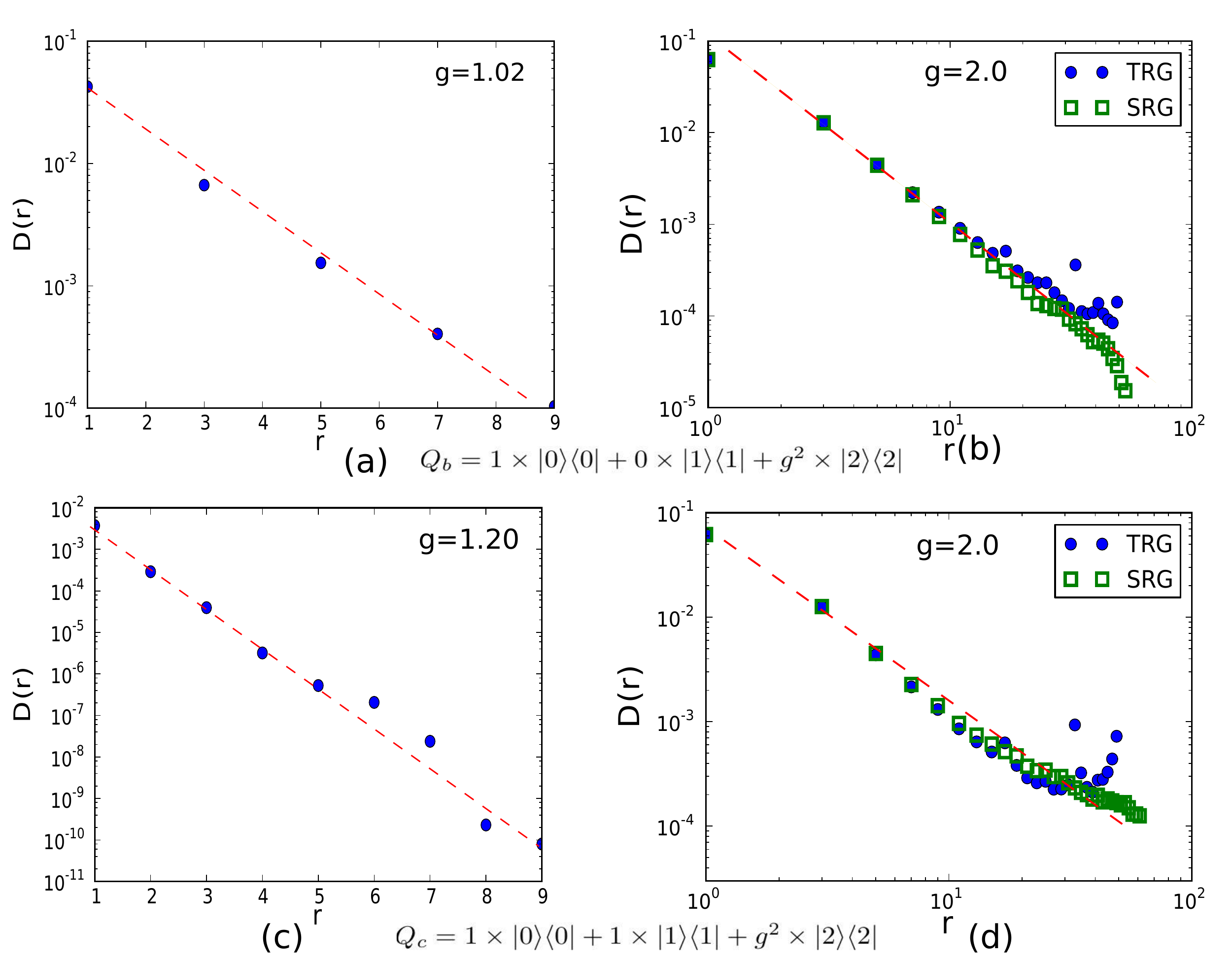,angle=0,width=9cm,height=10cm}}
\caption[] {  The correlation function under the deformation  $Q_b= 1 \times  |0 \rangle   \langle 0  |  +0 \times | 1  \rangle   \langle 1 |  +g^2 \times |2 \rangle   \langle 2 | $ with (a) $g=1.02$  (b) $g=2.0$. The correlation function under the deformation  $Q_c= 1 \times  |0 \rangle   \langle 0  |  + 1 \times | 1 \rangle   \langle 1 |  +g^2 \times |2 \rangle   \langle 2 | $ with (c) $g=1.3$  (d) $g=2.0$. } \label{Dr_all}
\end{figure}

\begin{figure*}[ht]
\center{\epsfig{figure=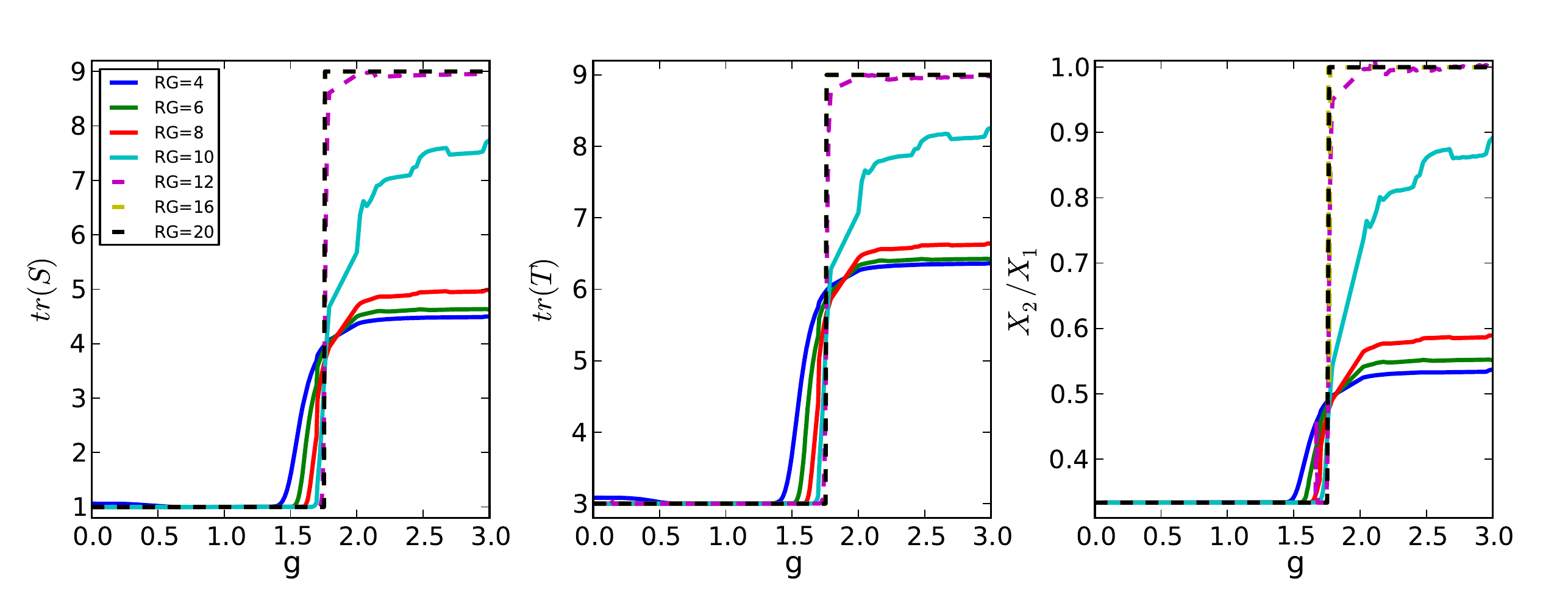,angle=0,width=17cm,height=6cm}}
\caption[]
{The trace of modular matrices  (a)$S$, (b) $T$, and the property  (c) $X_2/X_1$ as functions of parameter $g$ display a phase transition at critical point $g_c$  of  $\mathbb{Z}_3$   model with projector  $Q_c= 1 \times |0 \rangle   \langle 0  | +1 \times  |1 \rangle   \langle 1  |  +g^2 \times |2 \rangle   \langle 2 | $. }
   \label{11g_phases_giagram}
\end{figure*}

The GSPTRG can be used to detect the topological order well, but it cannot use to classify the trivial phases. 
It must look more carefully in trivial phases.
We need sufficient evidence to show what trivial phase is. 
The relevant quantity of our interest is the correlation function which appears useful in characterizing the product state and  critical phase. 
The connected correlation function is expressed as, 
\begin{align}
D(r) = \langle  S_0 (\vec{r}_i) S_0 (\vec{r}_i) \rangle -  \langle  S_0 (\vec{r}_j) \rangle  \langle  S_0 (\vec{r}_j) \rangle, 
\end{align} 
where $r= |\vec{r}_i - \vec{r}_j  |$ and $S_0 (\vec{r}_i)$ count 1 if $ ''0''$ state is present on the link $\vec{r}_i$ in a given configuration otherwise 
 $0$.
 In the asymptotic limit of large $r$, the correlation function will converge to zero.

When  $0 \leq g < 0.944$, we obtain topological entanglement entropy  $\gamma =0$.
In the limit $g=0$, this is a product state of all $0$.
There is good evidence to show that as $0 \leq g < 0.944$ the state in this region is the  trivial state.
When  $ g > 1.238$,  based on the observation of tensor representation  on $ g \gg 1$ discussed earlier, we would claim that this phase might be a critical phase. 
From the GSPTRG results, the values of $tr(S)$,  $tr(T)$, and $X_2/X_1$ approach to  $9.0$, $9.0$, and $1.0$ respectively. 
In the tensor renormalization approach, the above quantities are  calculated on  the square lattice with size $128 \times 128 $. 
 For the critical phase, the correlation function is algebraic of the form $D(r) \sim r ^{-b}$ for large $r$ as displayed in Fig.~\ref{Dr_all} (b) obtained using the tensor renormalization group  (TRG) and the mean-field approximated second renormalization group (SRG) \cite{SRG}.  
However, for TRG, the convergence becomes very slow and unstable in the critical phase~\cite{Huang2014}.
Even though the TRG is not efficient, the results with large critical bond dimension $\chi_c =64$ already show that the correlation function structure is different exponentially decaying  as shown in Fig.~\ref{Dr_all} (a). 
On the other hand, the mean-field approach of SRG can improve the accuracy of the results as shown in Fig.~\ref{Dr_all} (b) with bond dimension $\chi_c =64$. 
When  $0.994 \leq g < 1.238$, it is worth noting that the  $\mathbb{Z}_3$ phase exists, even if one of physical term $|1 \rangle$ is turned off. 
It is an unsettled question.

\subsection {For $ q_0=1; q_1=1; q_2=g^2$ }

Here we consider the deformation parametrized by $ q_0=1, q_1=1, q_2=g^2$ in the projector $Q_c= q_0 \times |0 \rangle   \langle 0  | +q_1 \times  |1 \rangle   \langle 1  |  +q_2 \times  |2 \rangle   \langle 2 | $.  The double tensor with nonzero components  of are given by
\begin{align}
\label{Z3quantum}
&  \mathbb{T}_{0000}^{0000} =  1  \\
&  \mathbb{T}_{1110}^{1110}=  \mathbb{T}_{1101}^{1101}=  \mathbb{T}_{1011}^{1011}=  \mathbb{T}_{0111}^{0111}=  1   \notag \\
& \mathbb{T}_{2220}^{2220}= \mathbb{T}_{2202}^{2202}= \mathbb{T}_{2022}^{2022}= \mathbb{T}_{0222}^{0222} =  g^6     \notag \\
& \mathbb{T}_{0012}^{0012} =   \mathbb{T}_{0120}^{0120} =   \mathbb{T}_{1200}^{1200} =  \mathbb{T}_{2001}^{2001} =  \mathbb{T}_{1002}^{1002} =   \mathbb{T}_{0021}^{0021}      \notag \\
& =\mathbb{T}_{0210}^{0210} =  \mathbb{T}_{2100}^{2100} =  \mathbb{T}_{0102}^{0102} =  \mathbb{T}_{1020}^{1020} =  \mathbb{T}_{0201}^{0201} =   \mathbb{T}_{2010}^{2010} =  g^2     \notag \\
&\mathbb{T}_{1122}^{1122} =  \mathbb{T}_{1221}^{1221} =  \mathbb{T}_{2211}^{2211} = \mathbb{T}_{2112}^{2112} =  \mathbb{T}_{1212}^{1212} =  \mathbb{T}_{2121}^{2121} =  g^4  \notag.
\end{align} 

At $g=1$, this is exactly  the $\mathbb{Z}_3$ phase and the corresponding state has topological order.
Recall the earlier case,  the same may be said,  while $g \gg 1$, the tensor form is also mathematically equivalent to the quantum dimer model at Rokhsar-Kivelson (RK) point on the square lattice, the wavefunction at which is the equal weight superposition of all dimer configurations.
At some critical point in $g$, the phase transition will occur. 

With the GSPTRG procedure, we flow the initial double tensor  for arbitrary $g$ to symmetry preserved  fixed point tensor and calculate  the modular matrices and the $X_2/X_1$. 
The results  are  shown in Fig.~\ref{11g_phases_giagram}.  
As the number of GSPTRG steps increase, the transition from $\mathbb{Z}_3$ phase to a  critical phase  approaches to a step function at  $g_c= 1.76$. 
From the correlation function, we also see that it is  algebraically decaying as shown in Fig.~\ref{Dr_all} (d).


\subsection {For $ q_0=1; q_1=g^2; q_2=g^2$ }

In the earlier section, we only consider that $g$ goes from $0$ to $1$. 
However, in the large $g$ limit, the effective terms of  Eg. \ref{Z3quantum} are $\mathbb{T}_{1122}^{1122} =  \mathbb{T}_{1221}^{1221} =  \mathbb{T}_{2211}^{2211} = \mathbb{T}_{2112}^{2112} =  \mathbb{T}_{1212}^{1212} =  \mathbb{T}_{2121}^{2121} = g^8$. 
We can regard string-$2$ as vacuum state and string-$1$ as dimer state. 
It will obey the rule of fully packed loop model where each vertex is connected to two dimers.  
Just as  the RK point of the quantum dimer model on the square lattice, the wavefunction at the large $g$ limit corresponds to the equal weight superposition of all fully packed loop. 
The RK-like point of fully packed loop model is also a critical liquid state with algebraically decaying correlation functions \cite{FPL1997}.  

\begin{figure}[ht]
\center{\epsfig{figure=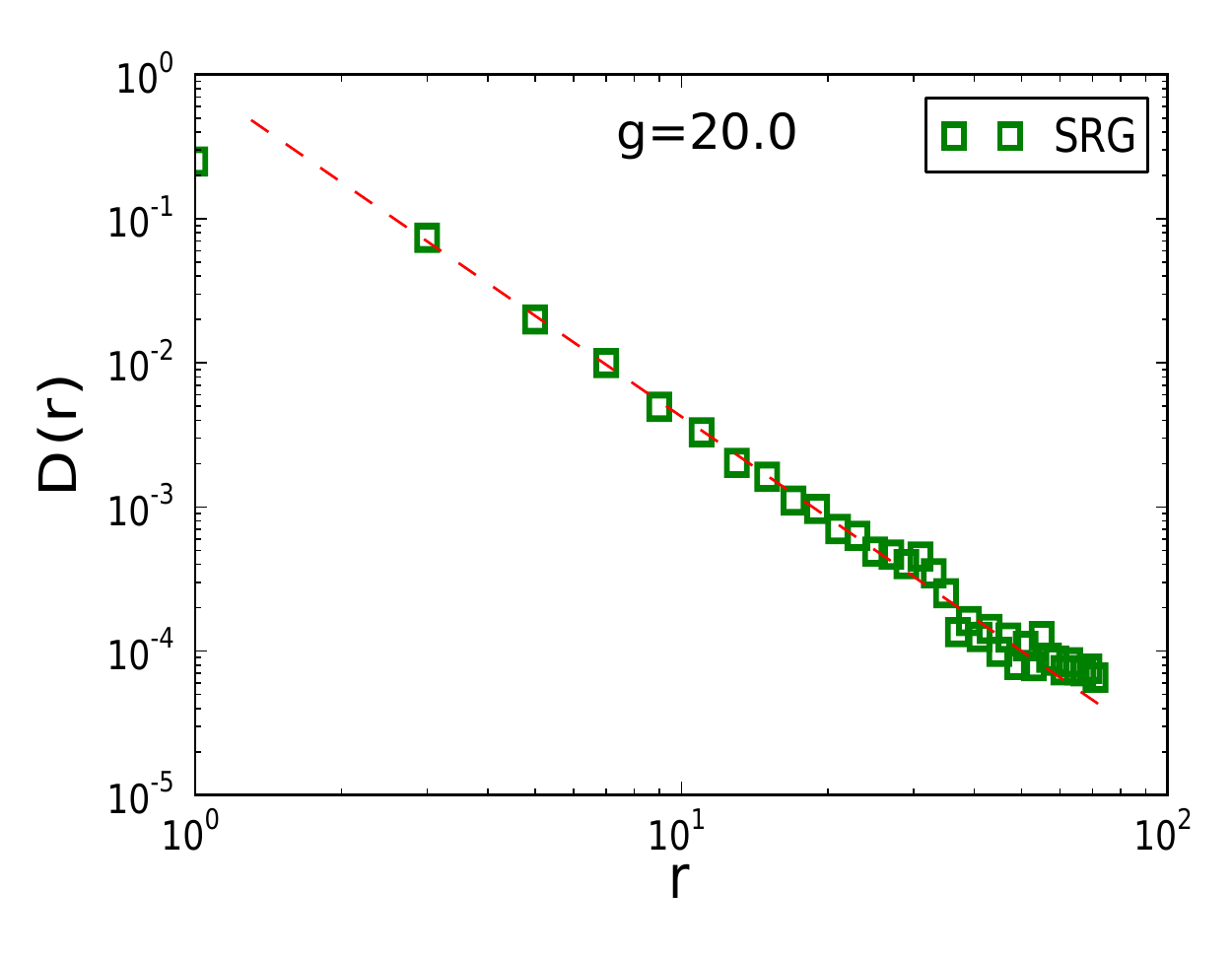,angle=0,width=8cm}}
\caption[]
{The correlation function under the deformation  $Q_a= 1 \times  |0 \rangle   \langle 0  |  + g^2 \times | 1  \rangle   \langle 1 |  +g^2 \times |2 \rangle   \langle 2 | $  with $g=20.0. $ } \label{Dr_1gg}
\end{figure}

The transition from $\mathbb{Z}_3$ phase to  the critical phase occurs at  $g_c= 7.3825$ from GSPTRG.  
We also calculate the correlation function by using SRG, and  see  the algebraically decaying beharior as shown in Fig.~\ref{Dr_1gg}.

The GSPTRG can be used to classify the topologically ordered phases with different modular matrices. 
The scheme is robust for topological order, because it is a gapped state and has gapped entanglement spectrum or singular value (SV) spectrum. 
When performing the RG transformation, as long as the cutoff in the bond dimension $\chi_{c}$  is large the wave function will flow to a fixed point.  
However, the scheme cannot  be used to distinguish different non-topological (trivial) phases.   
As we have observed, gapped and gapless trivial  phases under the GSPTRG scheme flow to fixed points that exhibit the same $tr(S)$, $tr(T)$, and $X_2/X_1$.
However,  this approach to fixed point is slower for the gapless phases as shown in Fig. \ref{10g_phases_giagram}.

\begin{figure}[ht]
\center{\epsfig{figure=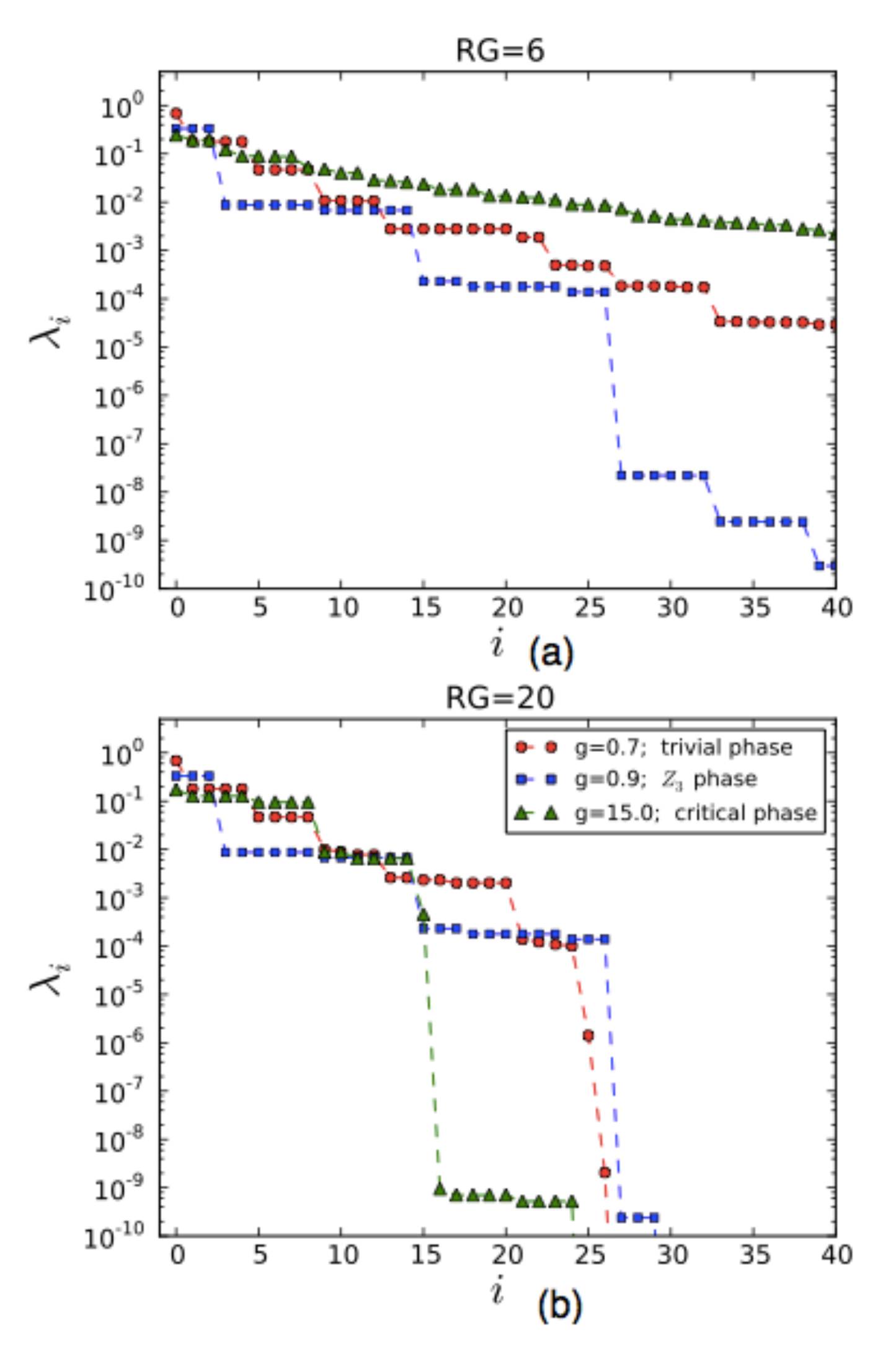,angle=0,width=9cm}}
\caption[]
{ Spectra of singular values of g=0.7, g=0.9 and g=15.0  for cutoff $\chi_c = 24$ after (a) $6$ and (b) $20$ applications of GSPTRG,  under the deformation  $Q_a= 1 \times  |0 \rangle   \langle 0  |  + g^2 \times | 1  \rangle   \langle 1 |  +g^2 \times |2 \rangle   \langle 2 | $. The index $i$ is to number the singular values $\lambda_i'$s.}
   \label{lambda_i}
\end{figure}

In Fig. \ref{lambda_i}, we plot the spectra of singular values after several steps of  GSPTRG transformation.
For $g=0.9$ which is a topological phase,   the there-fold degeneracy of  SV spectra can be found  and the pattern of degeneracy is robust under RG flow.  
The single largest SV separated by gap can be found in the gapped trivial phase.  
However, due to the finite cutoff $\chi_c$, the RG flow for the gapless phase cannot be followed accurately and indefinitely.
The continuous spectrum of SV eventually breaks into chunks separated by a gap after sufficiently large number of RG step; see Fig. \ref{lambda_i} (b). 
%
Increasing the bond dimension $\chi_{c}$ can slow the breakdown. 
An alternative, and perhaps the most elegant approach is proposed by Evenbly and Vidal \cite{TNR_2014}, who  proposed a coarse-graining transformation called tenser network renormalization (TNR) that can explicitly recovers the scale invariance and flow the wavefunction to a fixed point for critical points.
However, implementing the TNR scheme is beyond the scope of the present work.


\subsection{Summary of phase diagram}

\begin{figure}[ht]
\center{\epsfig{figure=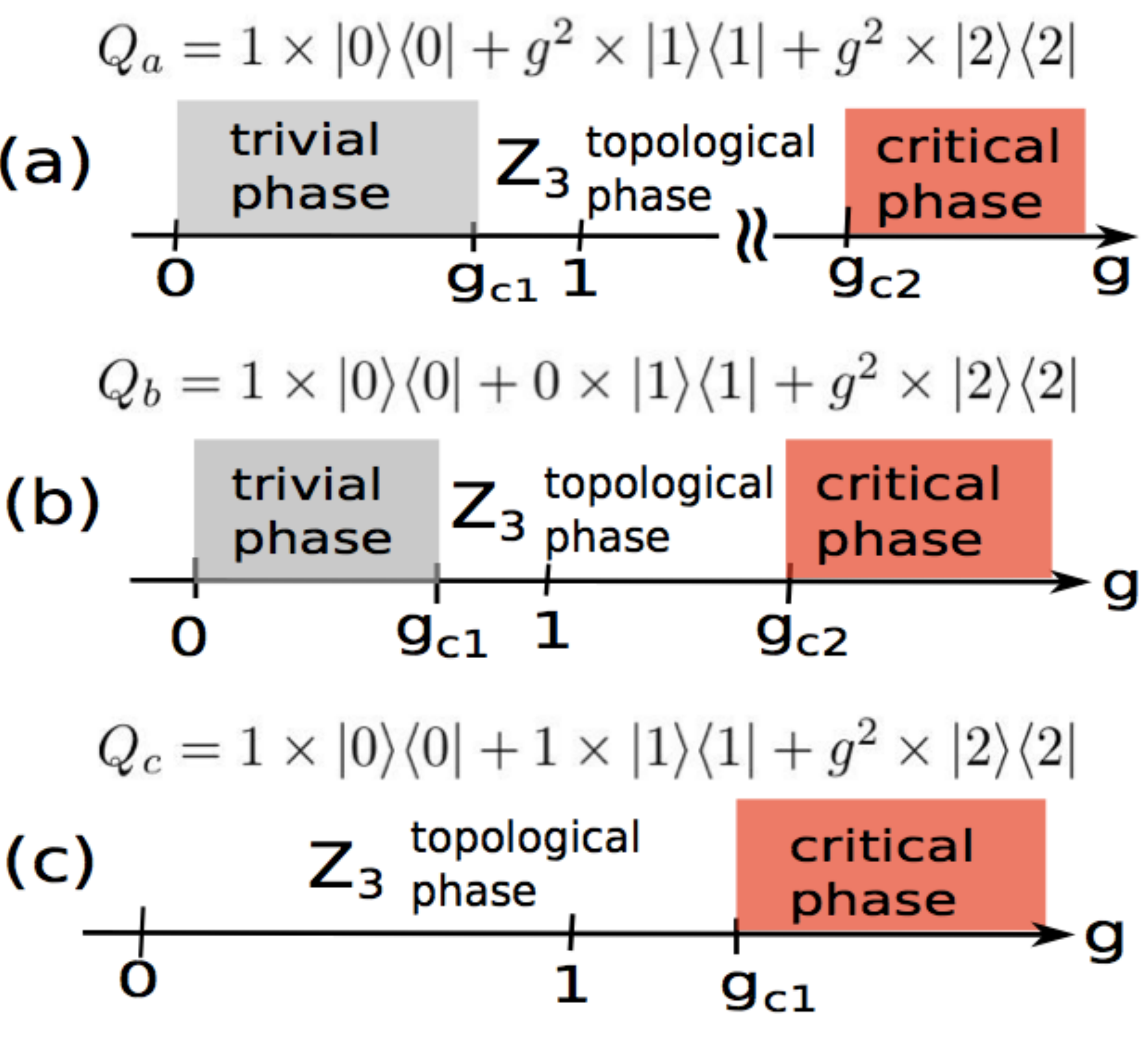,angle=0,width=8cm}}
\caption[]
{ Generic phase diagrams.  (a) the deformations  $Q_a= 1\times |0 \rangle   \langle 0  | + g^2 \times  |1 \rangle   \langle 1  |  +g^2 \times |2 \rangle   \langle 2 |   $,  (b) $Q_b= 1 \times  |0 \rangle   \langle 0  |  +0 \times | 1  \rangle   \langle 1 |  +g^2 \times |2 \rangle   \langle 2 | $,  (c) $Q_c= 1 \times|0 \rangle   \langle 0  | + 1\times  |1 \rangle   \langle 1  |  +g^2 \times |2 \rangle   \langle 2 |.  $ } \label{phase_diagram}
\end{figure}

The results of the section suggest the generic  $\mathbb{Z}_3$ model with projector $Q = q_0 \times |0 \rangle   \langle 0  | + q_1\times |1 \rangle   \langle 1  |  +q_2 \times |2 \rangle   \langle 2 |   $  shown in Fig.~\ref{phase_diagram}.
For projector $Q_a$, by varying parameter $g$, the phase transition from trivial state to topologically ordered phase will occur at $g_{c1}=0.7776$. 
The second phase transition  from trivial state to critical phase occur at $g_{c2}= 7.3825$. 
For projector $Q_b$, by varying parameter $g$, the first phase transition from trivial state to topologically ordered phase will occur at $g_{c1}=0.944$.
The second critical point  from  topologically ordered phase to critical phase is at $g_{c2}=1.238$.
The interesting thing is  that even we turn off one parameter, the $\mathbb{Z}_3$ phase can be found with two body system.
For projector $Q_c$, by varying parameter $g$, the phase transition from topologically ordered phase to critical phase  will occur at $g_{c1}= 1.76$. 
In general, richer phase diagrams may be obtained by considering more general parameter $q_0$, $q_1$, and $q_2$  or  $\mathbb{Z}_N$ model.

\subsection{For $N>3$ case}

In general, for other $\mathbb{Z}_N$  phases, we always can find the phase transition with deformation $Q=  \sum_{i=0} ^{N-1} q_i |i \rangle   \langle i | $. 
From tensor structure, sometimes we can deduce the possible phase. 
For example, we consider the $\mathbb{Z}_4$  phase with deformation  $Q= |0 \rangle   \langle 0 | + |1 \rangle   \langle 1 | + |2 \rangle   \langle 2| + g^2\times |3 \rangle   \langle 3 |$.  
As  $ g \gg 1$, the tensor from represent a product state of all $3$. 
This is  because  the $T_{3,3,3,3}$ was allowed. 
However, for the $\mathbb{Z}_3$  phase, the $T_{2,2,2,2}$ was not allowed. 
Thus, it is impossible to find the product state for  $\mathbb{Z}_3$  phase with $Q= |0 \rangle   \langle 0 | + |1 \rangle   \langle 1 | +  g^2 \times |2 \rangle   \langle 2 |$.

\section {Concluions} \label{conclusion}

We have employed the gauge symmetry protected tensor renormalization group (GSPTRG) method introduced by He, Moradi and Wen~Ref. \cite{He_Wen_2014} to study $\mathbb{Z}_N$ topological order under deformation.   
It is important to know the underlying gauge symmetry operators in order for this method to work. 
Due to the removal of irrelevant short-range entanglement, the fixed-point wavefunction  contains primarily the long-range entanglements that can be used to identify the topological order.
From the fixed-point form of the tensor representing the ground-state wavefunction, the modular matrices $S$ and $T$ that represent the mutual and self-statistics, respectively, of quasiparticles can be obtained. 
These can be used as order parameters to detect phase transitions. 
We applied the string tension to deform the $\mathbb{Z}_N$ wavefunction, similar to that in the $\mathbb{Z}_2$ toric code with deformation.   
The GSPTRG approach accurately determined the phase transition between the nontrivial topologically ordered phase and the trivial phase, the result of which matches very well with the mapping to $N$-state Potts model.
The RG process is a coarse-graining process and can be associated with the change of a length scale. 
From this perspective we were able to collapse the data from modular matrices with suitable scaling near the transitions and determined the critical exponent of the correlation length and the result agrees well with the mapping to the Potts model. 
In particular, we also investigated different deformations on the $\mathbb{Z}_3$ model and we found that the topologically ordered phase can be driven to  a critical phase.   Moreover, there exists a finite region of parameters such that the $\mathbb{Z}_3$ phase is composed of local two-level systems, i.e., qubits.

There has been tremendous progress on both theoretical and experimental advancement of the search of exotic phases with topological order.  
A future extension and application of the GSPTRG method would be to first use a gauge symmetry preserved approach to find the ground-state wavefunction of a Hamiltonian (potentially containing topological order) and use the RG to flow the wavefunction to a fixed point and obtain the modular matrices.

\section*{Acknowledgements}

The authors would like to thank Oliver Buerschaper, Lukasz Fidkowski,  Artur Garcia-Saez, Rom\'an Or\'us   for useful discussions.
This work was supported by the National Science Foundation under
Grants No. PHY 1314748 and No. PHY 1333903.


%

 \end{document}